\documentclass[12pt]{article}
\pdfoutput=1
\usepackage[width=420pt, height=640pt, vcentering]{geometry}
\usepackage[colorlinks=true]{hyperref}
\usepackage{amsmath}
\usepackage{amssymb}
\usepackage{amsthm}
\usepackage{graphics,graphicx}
\usepackage{lscape}
\usepackage[title,titletoc]{appendix}
\usepackage{footmisc}
\usepackage{rotating}
\usepackage{natbib}
\usepackage{enumitem}
\usepackage{textcomp}

\usepackage{bbm}

\newcommand{\ud}{\mathrm{d}}

\newcommand{\EE}{\mathbb{E}}

\newcommand{\tr}{^{\!\top}}

\title{ \textbf{Parallel American Monte Carlo}}
\author{\textbf{Calypso Herrera}\thanks{Quantitative Analyst, calypso.herrera@misys.com}  ~and  \textbf{Louis Paulot}\thanks{Head of Quantitative Research, Sophis, louis.paulot@misys.com} 
\\
[0.5cm]
{\it{Misys}\thanks{Sophis Quantitative Research, 42-44 rue Washington, 75008 Paris, France}}
}
\date{February 2014}

\begin{document}
\maketitle
\begin{abstract} 
In this paper we introduce a new algorithm for American Monte Carlo that can be used either for American-style options, callable structured products or for computing counterparty credit risk (\emph{e.g.} CVA or PFE computation). Leveraging least squares regressions, the main novel feature of our algorithm is that it can be fully parallelized. Moreover, there is no need to store the paths and the payoff computation can be done forwards: this allows to price structured products with complex path and exercise dependencies.
The key idea of our algorithm is to split the set of paths in several subsets which are used iteratively. We give the convergence rate of the algorithm. We illustrate our method on an American put option and compare the results with the Longstaff-Schwartz algorithm.
\end{abstract}

\section{Introduction}
American-style derivatives are found in all major financial markets. Monte Carlo simulation is used instead of the finite difference method when the products have more than two risk factors or have path dependencies. 

American Monte Carlo is also important in the context of CVA and PFE computations, where conditional expected values have to be computed at different times on simulation paths \citep{cesari2009modelling}.

The main disadvantage of Monte Carlo simulation is the computation time which is significantly higher than for a finite difference or trinomial method. This problem can easily be solved for European-style derivatives: both path generation and payoff computation can be parallelized and only the sum needs to be aggregated at the end. But this is not as simple for American options, callable structured products or CVA and PFE computations. 

The algorithm which is mainly adopted for its simplicity and its robustness is the Least Squares Monte Carlo (LSM) developed by \cite{longstaff2001valuing}. The American option is approximated by a Bermudan option. Starting from the final maturity, at each exercise date one compares the payoff from immediate exercise and the expected discounted payoff from continuation. Comparing the two values, one makes the decision to exercise or to hold the option. The conditional expectation is estimated from the information of all paths using a least squares regression.
However, this LSM algorithm with a backward recursion for approximating the price and the optimal exercise policy cannot be fully parallelized. Indeed, at each exercise date, the regression of the continuation value uses information from all paths, whose only generation can be parallelized. However and as opposed to European-style options, all paths must be kept in memory and sent to a single computation unit: once the paths are assembled, the least squares regressions, the optimal exercise decisions and the payoff estimation must be done by backward recursion.

In this article we address this bottleneck by introducing a new algorithm for American Monte Carlo that can be fully parallelized and relies on least squares regression to determine the optimal exercise strategy like LSM algorithm.

Our algorithm has several interesting features.
Firstly, all the steps of the computation can be parallelized.
Secondly, there is no need to keep the paths in memory or transfer them when the computation is done on a grid.
Thirdly, on each path the exercise decision and the payoff computation can be performed forwards. This allows complex path dependencies, including dependency on exercise decisions.
Fourthly, the algorithm allows the use of a technique known as \emph{boosting} in machine learning in order to get a more precise estimation of the exercise boundary.


The basic idea is the following. Instead of simulating all paths in a first phase and perform a backward recursion on all paths together, the set of paths is split in several subsets which are used iteratively. At each iteration, the coefficients of the regression are estimated using the paths of the previous iterations. A key observation is that in the equation of the least square regression, the information needed to compute regression coefficients is encoded in two objects which are linear in the paths (a matrix and a vector). Therefore they can be accumulated on paths of successive iterations without keeping all paths in memory. Only the linear system inversion has to be done at the beginning at each iteration, which can also be parallelized.

We prove the convergence of the price and compute the asymptotic error, or equivalently the convergence rate.

We finally illustrate our method with the computation of an American put option on a single factor. We compare the results and the computation performance with the LSM algorithm.

Early contributions to the pricing of American options by simulation were made in \cite{bossaerts1989simulation} and \cite{tilley1993valuing}. Other important works include \cite{barraquand1995numerical}, \cite{raymar1997monte}, \cite{broadie1997pricing}, \cite{broadie2004stochastic}, \cite{broadie1997enhanced}, \cite{ibanez2004monte} and \cite{garcia2000monte}. The idea of computing the expectation value of continuation using a regression was developed by \cite{carriere1996valuation}, \cite{tsitsiklis2001regression} and \cite{longstaff2001valuing}. 

Several recent articles propose the parallelization of the American option pricing include \cite{toke2006monte} and \cite{doan2010parallel}. These articles are based on the stratification or parametrization techniques to approximate the transitional density function or the early exercise boundary of \cite{ibanez2004monte} and \cite{picazo2002american}. 
Recent articles which address partial parallelization of the LSM algorithm include \cite{choudhury2008optimizations} and \cite{abbas2009american}. Unlike these articles which study the parallelization of the different phases of the LSM algorithm (path simulation, regression and pricing), we do not parallelize different phases of the LSM algorithm but we propose an innovative algorithm which can be fully parallelized.

Convergence of the LSM algorithm was addressed in several articles including \cite{CLP2001Conv} and \cite{Stentoft2004Conv}. 

Section \ref{sec:LS} presents the Longstaff-Schwartz algorithm, including the least squares regression. Section \ref{sec:algo} describes our new algorithm. Section \ref{sec:results} provides numerical results on the pricing of a put option. Section \ref{sec:conclusion} summarizes the results. Proofs, in particular the convergence rate, are presented in appendices.

\section{Longstaff-Schwartz Algorithm}
\label{sec:LS}

An American-style derivative gives the possibility to the holder to exercise it before maturity. The holder can choose at any time until the maturity to exercise the option or to keep it and exercise it later. Bermudan options are similar but exercise can happen only on specific dates. In order to price them, the American options are approximated by Bermudan options with discrete exercise dates. We consider that the state of the system is described by a vector of state variables $X_t$. In the simplest case, it is the spot value of the underlying asset. We assume that there exists a risk-neutral probability.
\subsection{Notations}
 We denote by $t_0$ the current date. Let us consider an option of maturity $T$ and $N$ early exercise dates. 

We will use the following notations:
\begin{itemize}
\item $T$ maturity of the option
\item $t_0$ computation date
\item $t_0 < t_1 < . . . < t_{M-1} < t_M = T$ discretized exercise dates
\item $X_t$ vector of state variables
\item $X_k = X_{t_k}$ value of the state variable vector at date $t_k$
\item $C_k = C_k(X_k)$ continuation value at date $t_k$
\item $\widehat{C}_k = \widehat{C}_k(X_k)$ approximation of the continuation value at date $t_k$
\item $F_k = F_k(X_k)$ payoff value in case of exercise at date $t_k$
\item $P_k = P_k(X_k)$ discounted value of the option, with optimal exercise at date $t_k$ or later.
\item $r_t$ instantaneous interest rate at time $t$
\end{itemize}

\subsection{Least squares regression}
\label{sec:LeastSquare}

The American option can be valued using the following recursion.
At option maturity, the value of the option is equal to the payoff value $P_M = F_M$.
At a previous date $t_k$ the holder has two possibilities:
\begin{itemize}[topsep=2pt,itemsep=4pt,parsep=0pt]
\item[$\bullet$] exercise the option and get the cashflow $F_k$;
\item[$\bullet$] keep the option at least until the next exercise time $t_{k+1}$. If we assume there is no arbitrage opportunity, the continuation value of the option is the expected discounted value of the option, conditionally to the information available at time $t_k$:
\begin{equation}
C_k= \EE\Big(e^{-\int_{t_k}^{t_{k+1}}r_s ds} P_{k+1} \Big| X_k\Big) \rlap{\ .}
\label{eq:CondExpect}
\end{equation}
\end{itemize}
The holder will exercise if the payoff $F_k$ is higher than the continuation value $C_k$. Therefore, at time $t_k$ the discounted optimally exercised payoff is
\begin{equation*}
P_k = \left\{ \begin{array}{ll}F_k & F_k \geq C_k \\
e^{-\int_{t_k}^{t_{k+1}}r_s ds} P_{k+1} & F_k < C_k \rlap{\ .}
\end{array} \right.
\end{equation*}

In a Monte Carlo computation, the conditional value in \eqref{eq:CondExpect} is not trivially available. One way to estimate it is to approximate it as a linear combination of basis functions\footnote{This finite linear expansion can be seen as the projection of the infinite-dimensional functional space on a finite-dimensional subspace, or equivalently as the truncation of a linear expansion on an infinite number of Hilbert basis functions. There are several choices of the basis functions, giving different qualities of approximation.}:
\begin{equation}
C_k(X_k) = \EE\big(\widetilde{P}_{k+1} \big| X_k\big) \simeq  \widehat{C}_k(X_k) = \sum_{l=1}^p \alpha_{k,l} f_{k,l}(X_k)
\label{eq:ContinuationExpansion}
\end{equation}
with
$$\widetilde{P}_{k+1} = e^{-\int_{t_k}^{t_{k+1}}r_s ds} P_{k+1} \rlap{\ .}$$

Coefficients $\alpha_{k,l}$ in \eqref{eq:ContinuationExpansion} are estimated using the least squares method. In other words, they are chosen to minimize a quadratic error function. Denoting by $\alpha_k$ the vector of coefficients, for each date $t_k$ we want to minimize
\begin{equation*}
\Psi_k(\alpha_{k}) = \EE\left[w_k(X_k) \left(C_k - \sum_{l=1}^p \alpha_{k,l} f_{k,l}(X_k) \right)^2\right] \rlap{\ .}
\end{equation*}
where $w_k(X_k)$ are weights which allow to give a different weight to each path. The choice of Longstaff and Schwartz is to take the weight equal to 1 when the option is in the money at time $t_k$ and 0 otherwise.

$C_k$ is the conditional expected value $\EE\big(\widetilde{P}_{k+1} \big| X_k\big)$ and is not known, as it is the function we want to estimate. Therefore the least square regression cannot be directly applied. However, minimizing $\Psi_k$ is equivalent to minimizing a different function:
\begin{equation*}
\Phi_k(\alpha_{k}) = \EE\Bigg[w_k(X_k) \bigg(\widetilde{P}_{k+1} - \sum_{l=1}^p \alpha_{k,l} f_{k,l}(X_k) \bigg)^2\Bigg] \rlap{\ .}
\end{equation*}
See appendix \ref{app:continuation} for a proof. The difference between $\Phi_k$ and $\Psi_k$ is that there are no more conditional expectations.
Thus the coefficients of the basis functions can be estimated using the least square method, by regressing the discounted option values $\widetilde{P}_{k+1}$ on the state variable values $X_k$ at $t_k$.

In practice, the expected value in $\Phi_k$ which is minimized is, up to an irrelevant factor $N$, the Monte Carlo estimation
\begin{equation*}
\Phi_k(\alpha_{k,l}) = \sum_{j=1}^N w_k\left(X_k^{(j)}\right) \left(\widetilde{P}_{k+1}^{(j)} - \sum_{l=1}^p \alpha_{k,l} f_{k,l} \left(X_k^{(j)}\right) \right)^2
\end{equation*}
where $X_k^{(j)}$ is the state variable vector on path $j$ at time $t_k$ and $\widetilde{P}_{k+1}^{(j)}$ is the value of the stochastic variable $\widetilde{P}_{k+1}$ on path $j$. The weights $w_k\left(X_k^{(j)}\right)$ allow to focus on the more relevant paths, as explained in section \ref{sec:weights}.

This function $\Phi_{k}$ has a minimum on $\alpha_k$ when the partial derivative with respect to $\alpha_{k,l}$ are zero for all $l$:
\begin{multline}
\frac{\partial\Phi_{k}}{\partial \alpha_{k,l}} = 
2  \sum_{m=1}^p \sum_{j=1}^N w_k\Big(X_k^{(j)}\Big) f_{k,l}\Big(X_k^{(j)}\Big) f_{k,m}\Big(X_k^{(j)}\Big) \alpha_{k,m}
\\
  - 2  \sum_{j=1}^N w_k\Big(X_k^{(j)}\Big) f_{k,l}\Big(X_k^{(j)}\Big) \widetilde{P}_{k+1}^{(j)} 
= 0 \ .
\label{eq:min}
\end{multline}
Let us introduce $p \times p$ matrix $U_k$ and dimension $p$ vector $V_k$
\begin{eqnarray*}
U_{k,lm} &=& \sum_{j=1}^N w_k\Big(X_k^{(j)}\Big) f_{k,l}\Big(X_k^{(j)}\Big) f_{k,m}\Big(X_k^{(j)}\Big)
\\
V_{k,l} &=& \sum_{j=1}^N w_k\Big(X_k^{(j)}\Big) f_{k,l}\Big(X_k^{(j)}\Big) \widetilde{P}_{k+1}^{(j)}
\end{eqnarray*}
or in a simpler, vectorial notation
\begin{eqnarray}
U_k &=& \sum_{j=1}^N w_k\Big(X_k^{(j)}\Big) f_k\Big(X_k^{(j)}\Big) f_k^\top\!\Big(X_k^{(j)}\Big)
\nonumber
\\
V_k &=& \sum_{j=1}^N w_k\Big(X_k^{(j)}\Big) f_k\Big(X_k^{(j)}\Big) \widetilde{P}_{k+1}^{(j)}
\rlap{\ .}
\label{eq:UV-LS}
\end{eqnarray}
We can rewrite equation \eqref{eq:min} as
$$
U_k \alpha_k = V_k \rlap{\ .}
$$
For each date $t_k$ coefficients $\alpha_k$ are therefore obtained through matrix inversion or using a linear equation solver:
$$
\alpha_k = U_k^{-1} V_k \rlap{\ .}
$$
This is the vector of coefficients which minimizes the quadratic error function. It gives the least square estimation of the continuation value of the option at time $t_k$ on $N$ Monte Carlo paths:
\begin{equation*}
\widehat{C}_k(X_k)  = \sum_{l=1}^p \alpha_{k,l} f_{k,l}(X_k) = \alpha_k^\top f_{k}(X_k) \rlap{\ .}
\end{equation*}

When the coefficients are estimated, they are used to compute the continuation value at time $t_k$ for each path. The continuation value will be used for the decision to continue or to exercise the option. When the decision is made,  we have the cashflow at time $t_k$. If the decision is to continue, we use the simulated value of the payoff and not the estimated value. For each path, we compute the cashflow for all dates backwards. 

\subsection{Algorithm}

In summary, the Longstaff-Schwartz algorithm is the following:
\begin{enumerate}
\item Simulate $N$ Monte Carlo paths $X_k^{(j)}$ ($1 \leq j \leq N$, $1 \leq k \leq M$) and keep them in memory or store them.
\item On the last date $t_M$, compute the terminal payoff $P_M^{(j)} = F_M\Big(X_N^{(j)}\Big)$ on all paths $j$.
\item Starting from $k=M-1$ and until $k=1$, perform a backward recursion:
\begin{enumerate}
\item Summing over all paths and using payoff value at date $t_{k+1}$, compute
\begin{eqnarray*}
U_k &=& \sum_{j=1}^N w_k\Big(X_k^{(j)}\Big) f_k\Big(X_k^{(j)}\Big) f_k^\top\!\Big(X_k^{(j)}\Big)
\\
V_k &=& \sum_{j=1}^N w_k\Big(X_k^{(j)}\Big) f_k\Big(X_k^{(j)}\Big) \widetilde{P}_{k+1}^{(j)}
\rlap{\ .}
\end{eqnarray*}
\item Get least square coefficients
$
\alpha_k = U_k^{-1} V_k \rlap{\ .}
$
\item On every path $j$, compare the payoff value $F_k\Big(X_k^{(j)}\Big)$ and the continuation value estimate
$\widehat{C}_k\Big(X_k^{(j)}\Big)  =  \alpha_k^\top f_{k}\Big(X_k^{(j)}\Big)$. If $ F_k\Big( X_k^{(j)}\Big) \geq \widehat{C}_k\Big(X_k^{(j)}\Big)$, set $P_k^{(j)} = F_k\Big( X_k^{(j)}\Big)$; else set $P_k^{(j)} = \widetilde{P}_{k+1}^{(j)} = e^{-\int_{t_k}^{t_{k+1}} r_s \ud s} P_{k+1}^{(j)}$.
\end{enumerate}
\item Finally get the Monte Carlo estimate of the derivative price as
\begin{equation}
P = \frac{1}{N} \sum_{j=1}^N P_1^{(j)} \rlap{\ .}
\label{eq:Price-LS}
\end{equation}
\end{enumerate}

\subsection{Limitations}

The Longstaff-Schwarz algorithm is powerful and allows to price multi-factor, path-dependent derivatives with early exercise using Monte Carlo simulations. However, we can state a few limitations
\begin{description}
\item[Parallelization]
Monte Carlo pricing is time-consuming. In order to get good performance, we want to parallelize the computations. In the standard American Monte Carlo algorithms, such as the Longstaff-Schwartz algorithm that we described, only the path generation can be parallelized. Since it makes use of all paths, the backward regression has to be done on a single computation unit. This includes the least square estimation of the continuation value, the exercise decision and the computation of $P_k^{(j)}$ on each path.
\item[Memory consumption] Since all paths must be generated in a first phase and used in a second one, all paths must be stored. For an option with several underlyings and many exercise dates, this can represent large amounts of data. In addition, if the path generation is distributed on some grid, it means that a large quantity of data must be transferred.
\item[Limited path dependence] As the payoff is computed backwards, the path dependence is limited to quantities present in the state variables vector. It can include quantities which depend on past values on a given path but not quantities which depend on exercise decisions at previous dates.

As an example, a swing option allows to buy some asset (usually electricity or gas) at several dates for a price fixed in the contract, with some global minimum and maximum on the total quantity. This means that exercising on date $t_k$ depends on the exercises on date $t_{k'}$, $k' <k$. This cannot be directly handled by a standard Longstaff-Schwartz algorithm.
\end{description}

\section{Parallel iterative algorithm}
\label{sec:algo}

We propose an algorithm for American Monte Carlo with the following properties.

\begin{description}
\item[Full parallelization] All phases of the computation can be parallelized.
\item[No path storage] Monte Carlo paths are used only once. There is no need to keep them in memory or transfer them when the computation is done on a grid. Only fewer aggregated data are kept in memory and exchanged between computation units.
\item[Forward computation] On every path, exercise decisions and payoff computation can be performed forwards from $t_1$ to $t_M$. This allows all kinds of path-dependence, including dependence on previous exercise decisions.
\item[Boosting] The algorithm allows to use some \emph{boosting} in order to get more and more precise estimates of the exercise boundaries.
\item[More general regression] Least square regression can be performed for all or several dates together, introducing exercise time as a variable of the continuation value function.
\end{description}

\subsection{Iterations}

Instead of simulating all paths in a first step and performing a backward recursion on all paths together in a second step, the $N$ paths are split in several sets which are used iteratively. On each iteration, coefficients $\alpha_k$ are estimated using paths of the previous iterations. A key observation is that in equation \eqref{eq:UV-LS} $U_k$ and $V_k$ are linear in the paths. The information needed to compute regression coefficients is encoded in these objects and can be accumulated on paths and successive iterations without keeping all paths in memory. Only the linear system inversion has to be done at the beginning of each iteration.

For a given iteration, the exercise decisions depend on objects $U_k$ and $V_k$ obtained in previous iterations. Within this iteration, computations on different paths are independent from each other. This means that they can be run in parallel. Once quantities from all paths in a given iteration are accumulated, solving the linear system can be done independently for every date. Therefore this can also be parallelized.

In addition, as the exercise decision is made using information from previous iterations, there is no need to use a backward computation: all payoff computations and exercise decisions can be done in the natural order. (Note that in simple cases, it may however require less calculations to do it backwards on a given path.)

One may think that using only a limited number of paths to make the exercise decisions in the first iterations will increase the error in the final price. However it appears that this effect is small after a few iterations. In order to reduce the error in the final results, we introduce weights depending on the iteration in both formulas \eqref{eq:UV-LS} and \eqref{eq:Price-LS}: paths from first iterations are less weighted than paths from the following iterations which are more precise.

In fact, the iterative nature of our algorithm even allows to use something similar to what is called \emph{boosting} in machine learning, as already introduced in the context of American options pricing in \cite{picazo2002american}. This can eventually give smaller errors than classical Longstaff-Schwartz algorithm.

\subsection{Notations}

The $N$ paths are partitioned in $n$ distinct sets. Let us assume each piece of the partition is made of consecutive paths and denote by $n_i$, $1 \leq i \leq n$ the final path of each set. This means that the $i$th iteration will use paths from $n_{i-1}+1$ to $n_i$.

\begin{itemize}
\item $M$: number of exercise dates.
\item $N$: total number of paths.
\item $n$: number of iterations.
\item $n_i$: last path of $i$th iteration. Iteration $i$ uses path $n_{i-1}+1$ to $n_i$, with $n_0=0$ and $n_n=N$. 
\item $\widetilde{w}_i$: weight of paths of the $i$th iteration in the price sum.
\item $w_k^{(i)}\!\Big(X_k^{(j)}\Big)$: weight of path $j$ inside iteration $i$ in matrix $U$ and vector $V$ sums in equations \eqref{eq:UV-LS}. A special case is the factorization $w_k^{(i)}(X_k) = w_i y_k(X_k)$.
\item $U_k^{(i)}$ and $V_k^{(i)}$: matrices and vectors containing information from path 1 to $n_i$ and used to compute $\alpha_k^{(i)}$.
\item $u_k^{(i)}$ and $v_k^{(i)}$: contributions of iteration $i$ to $U_k^{(i)}$ and $V_k^{(i)}$, containing information from paths $n_{i-1}+1$ o $n_i$.
\item $\alpha_k^{(i)}$: vector of coefficients regressed on paths 1 to $n_i$.
\item $\widehat{C}^{(i)}_k(X_k)$: approximated continuation value given by coefficients $\alpha_k^{(i)}$. It is used in iteration $i+1$.
\item $\kappa_k^{(j)}$: optimal exercise time index $\kappa$, such that optimal exercise time is $t_\kappa$ on path $j$ if not exercised before $t_k$.
\item $P_k^{(j)}$: discounted payoff on path $j$ at time $t_k$ if the option is not exercised before.
\item $P^{(i)}$: sum of option discounted payoff from paths of the $i$th iteration.
\item $\bar P_{N}$: total weighted sum of discounted option payoff from paths 1 to N.
\item $q_{N}$: sum of price weights of paths 1 to N.
\end{itemize}

\subsection{Algorithm}

\begin{enumerate}

\item Initiate the algorithm using rough estimates for exercise boundaries or coefficients $\alpha_k^{(0)}$:
for example consider that the option is exercised at final maturity only or, alternatively, use final coefficients from previous day computation.
\item Iterate on $i$ from 1 to $n$:
\begin{enumerate}[leftmargin=10pt]
\item Iterate on $j$ from $n_{i-1}+1$ to $n_i$:
\begin{enumerate}[leftmargin=10pt]
\item Simulate path $j$ and get state variable vector $X_k^{(j)}$ at all dates.
\item 
For all dates $t_k$, compare the payoff value $F_k\Big(X_k^{(j)}\Big)$ and the continuation value estimate from the previous iteration
$$\widehat{C}_k^{(i-1)}\Big(X_k^{(j)}\Big)  = \alpha_k^{(i-1) \top} f_k\Big(X_k^{(j)}\Big) \rlap{\ .}$$
From this, for all $k$ get\footnote{This computation can be done in a forward manner, however numerically, the fastest way to perform this computation is to do it backwards. Starting on the last exercise date $t_M$ we set $P_M^{(j)} = F_M\Big(X_M^{(j)}\Big)$. Then recursively on $k$, if $F_k\Big( X_k^{(j)}\Big) \geq \widehat{C}_k^{(i-1)}\Big(X_k^{(j)}\Big)$, set $P_k^{(j)} = F_k\Big( X_k^{(j)}\Big)$; else set $P_k^{(j)} = \widetilde{P}_{k+1}^{(j)} = e^{-\int_{t_k}^{t_{k+1}} r_s \ud s} P_{k+1}^{(j)}$.}
$$
\kappa_k^{(j)} = \min\bigg( k' \geq k \ \bigg| \ k' = N
\text{\ or\ } F_{k'}\Big( X_{k'}^{(j)}\Big) \geq \widehat{C}_{k'}^{(i-1)}\Big(X_{k'}^{(j)}\Big)\bigg)
$$
and finally $P_k^{(j)} =e^{-\int_{t_k}^{\kappa_k^{(j)}} r_s \ud s} F_{\kappa_k^{(j)}}\Big(X_{\kappa_k^{(j)}}^{(j)}\Big) $ and $\widetilde{P}_{k+1}^{(j)} = e^{-\int_{t_k}^{t_{k+1}} r_s \ud s} P_{k+1}^{(j)}$.

\item Accumulate the contribution to the price 
$$
P^{(i)} = \sum_{j=n_{i-1}+1}^{n_i} P_1^{(j)}  \rlap{\ .}
$$
\item For every date $t_k$ add the contribution of path $j$ to
 \begin{eqnarray*}
u^{(i)}_k &=& \sum_{j=n_{i-1}+1}^{n_i} w_k^{(i)}\!\Big(X_k^{(j)}\Big) f_k\Big(X_k^{(j)}\Big) f_k^\top\!\Big(X_k^{(j)}\Big)
\\
v^{(i)}_k &=& \sum_{j=n_{i-1}+1}^{n_i} w_k^{(i)}\!\Big(X_k^{(j)}\Big) f_k\Big(X_k^{(j)}\Big) \widetilde{P}_{k+1}^{(j)}
\rlap{\ .}
\end{eqnarray*}
\end{enumerate}
\item For every date $t_k$, add the contributions $u^{(i)}_k$ and $v^{(i)}_k$ of iteration $i$ to\footnote{When the weight $w_k^{(i)}(X_k)$ factorizes as $w_k^{(i)}(X_k) = w_i y_k(X_k)$, the multiplication by $w_i$ can be factorized at this step: $u^{(i)}_k = \sum_{j=n_{i-1}+1}^{n_i} y_k\!\left(X_k^{(j)}\right) f_k\left(X_k^{(j)}\right) f_k^\top\!\left(X_k^{(j)}\right)$ and $U^{(i)}_k = U^{(i-1)}_k + w_i u^{(i)}_k$ and similarly for $V$.}
 \begin{eqnarray*}
U^{(i)}_k &=& \sum_{l=1}^i u^{(l)}_k
\\
V^{(i)}_k &=& \sum_{l=1}^i v^{(l)}_k
\rlap{\ ,}
\end{eqnarray*}
solve the linear system
$$
U_k^{(i)} \alpha_k^{(i)} = V_k^{(i)}
$$
and get the coefficients of the least squares regression on $n_i$ first paths:
$$
\alpha_k^{(i)} = \Big(U_k^{(i)}\Big)^{-1} V_k^{(i)}  \rlap{\ .}
$$
\item Using price weights $\tilde{w}_i$, accumulate the contributions of iteration $i$ to
\begin{eqnarray*}
\bar P_N &=& \sum_{i=1}^n \widetilde{w}_i P^{(i)}
\\
q_N &=& \sum_{i=1}^n \widetilde{w}_i (n_i-n_{i-1}) \rlap{\ .}
\end{eqnarray*}
\end{enumerate}
\item Finally get the Monte Carlo estimate of the option price as the weighted average
\begin{equation*}
P = \frac{ \bar P_N}{q_N} \rlap{\ .}
\end{equation*}
\end{enumerate}

\subsection{Parallel computing}

For every iteration, steps (a) and (b) can inherently be parallelized. In step (a), all the paths in a given iteration are independent from each other and computation related to different paths can be run in parallel. Similarly, the linear systems for different dates in (b) can be solved in parallel.

The data which must be shared or transfered between computation units are objects $U_k$ and $V_k$ for all dates, coefficients $\alpha_k$ and contribution to the final price $P^{(i)}$.

\subsection{Convergence}
\label{eq:convergence}

We assume weights $\widetilde{w}_i \sim 1$ when $i \rightarrow \infty$. We also assume that $w_k^{(i)}(X_k)$ factorizes as $w_k^{(i)}(X_k) = w_i y_k(X_k)$ with  $w_i \sim 1$ when $i \rightarrow \infty$.

Let us fix a vector of initial regression coefficients $\alpha$. Using these coefficients in exercise decisions, let us define $\bar u(\alpha) = \EE\big[f(X) f(X)^\top\big]$ and $\bar v(\alpha) = \EE\big[f(X) \widetilde{P}\big]$. This gives a function $\alpha \mapsto \bar \alpha(\alpha) = \bar u(\alpha)^{-1} \bar v(\alpha)$. This corresponds to the vector of coefficients obtained after a single iteration in the limit of an infinite number of paths.
Let us assume this function $\alpha \mapsto \bar \alpha(\alpha)$ is contractant, \emph{i.e.} Lipschitz-continuous
$$
\forall \alpha, \alpha' \quad \| \bar\alpha(\alpha) - \bar\alpha(\alpha') \| \leq q \| \alpha - \alpha' \|
$$
with\footnote{For an American option, the continuation value for a given date reaches a maximum when the estimated continuation value is exact for the following dates. As a consequence, $\frac{\partial\bar{\alpha}}{\partial \alpha}$ vanishes for the optimal $\alpha$. Around this point, it is not a strong constraint to assume that the function is contractant.} $q < 1$.

The Banach fixed-point theorem then ensures this function has a fixed point. Let us denote by $A$ the norm of the (matrix) operator $\frac{\partial \bar \alpha(\alpha)}{\partial \alpha}$ at this fixed point. We have $A \leq q < 1$.

Let us assume there are $n$ iterations of $m$ paths, with a total number of paths $N = nm$.

Then the algorithm we propose converges to an approximation of the price as $n \to \infty$.

As the continuation value is projected on a finite dimensional basis, exercise boundaries are approximations and therefore the exercise is slightly sub-optimal. As a consequence, the algorithm converges to a value which is lower than the real price. When the number of basis functions grows, the price estimate becomes closer to the real price. The same behavior is observed in Lonstaff-Schwartz algorithm. 
The error term around this limit value has an expected value in $O\left(\frac{1}{n^{1-A}}\right)$ and a standard error in $O\left(\frac{1}{\sqrt{m\, n^{\max(1,2-2A)}}}\right)$. If $A \leq \frac{1}{2}$, this is the usual Monte Carlo error $O\left( \frac{1}{\sqrt{nm}}\right) = O\left( \frac{1}{\sqrt{N}}\right)$.

When the weights of the paths in $y_k(X^k)$ are the same as chosen by Longstaff and Schwartz, 1 in the money and 0 out of the money, then the algorithm converges to the same price as Longstaff-Schwartz algorithm.

The proof is given in appendix \ref{app:convergence}.
\subsection{Path weights}
\label{sec:weights}
In order to improve the convergence of the algorithm, the paths can be given different weights, in the computation of matrix $U$ and vector $V$ on one hand, and in the price computation on the other hand.

\subsubsection{Exercise boundary}
Longstaff and Schwartz use a simple weight for paths in the regression: at date $t_k$, path $j$ is taken into account only if the option is in the money at date $t_k$. The weight $w_k(X_k^{(j)})$ is equal to 1 when the option is in the money and 0 otherwise. This is used for the computation of the matrix $U_k$ and the vector $V_k$ in the equation \eqref{eq:UV-LS}.
This weight improve the convergence of the algorithm: the paths in the money are the only paths eligible to be exercised.

Going further, we want to concentrate on paths which are closed to the exercise boundary. In addition, we require the weight to be continuous, which will give smoother greeks. 

In the case of a product on one underlying, we suggest a simple weight function:
\begin{equation*}
y_k(X_k) =e^{-\frac{\left(X_k - B_k\right)^2}{2\beta_k^2}}
\end{equation*}
where $X_k$ is the spot price at the date $t_k$ and $B_k$ is the exercise boundary value at the same date. At each date $t_k$, the exercise boundary is the solution of the equation 
$ F_k(x) = \widehat{C}_k(x)$ where $F_k$ is the payoff value and $ \widehat{C}_k$ is the continuation value estimate. The boundary is computed using the coefficients $\alpha_k$ of the previous iteration. This equation can be approximatively solved with a simple numerical method.

Parameters $\beta_k$ are chosen to give a good compromise between statistical error and systematic error. The statistical error is reduced for large $\beta_k$, when many paths are taken into account. The systematic error is reduced when we only look at paths close to the exercise boundary, for small $\beta_k$. We can use the iterative nature of our algorithm to reduce $\beta_k$ as the number of iterations grows. This would allow both statistical and systematic error to be reduced. This is similar to \emph{boosting} in machine learning: as the number of iterations increases, we concentrate more closely around the exercise boundary.

\subsubsection{Iterations and weights on $U$,$V$}
As the algorithm is iterative, the values of the regression coefficients are not precise in the first iterations. For this reason, a simple optimization of the algorithm is to give a low weight to the first iterations. At each iteration, the matrix $U_k^{(i)}$ and the vector $V_k^{(i)}$ are filled and added to the $U_k^{(i-1)}$ and $V_k^{(i-1)}$ of the previous iteration. We introduce a weight which increases with the number of iteration $i$: $w_i = \prod_{j=i+1}^n w_{UV}^{(i)}$ with
\begin{equation}
w_{UV}^{(i)} = 1 - \lambda e^{-\frac{i}{\mu}}  \rlap{\ .}
\label{eq:wUV}
\end{equation}
Each $U_k^{(i)}$ and $V_k^{(i)}$ from previous iteration are multiplied by $w_{UV}^{(i)}$. This decreases the weight of first iterations in the regression coefficients.

\subsubsection{Iterations and weights on price}

Similarly, during the first iterations the estimated continuation value is not accurate as the coefficients $\alpha_k$ are not and therefore neither the price. A simple way to improve the convergence is to eliminate the first paths from the computation of the final price. For this reason the final price is a weighted average where the first paths do not have an important weight. We introduce a weight $\widetilde{w}_i$ which depends on the iteration. The weight increases with the iterations. 

We use the following function:
\begin{equation}
\widetilde{w}_i = 1 - \frac{1}{2}\left(1-\tanh \left[\nu(i -1)\right]\right) \rlap{\ .}
\label{eq:wP}
\end{equation} 
At each iteration $i$, we multiply the sum of present values of iteration $i$ by this weight $\widetilde{w}_i$ before adding to the sum of present values of the previous iterations. 

\subsection{Time as a variable of regression functions}

Finally one can leverage the iterative nature of our algorithm to lower the total number of basis functions in the regression and decrease the statistical error of the least squares estimation.

In the simplest algorithm, the regression is made independently for each date: for each date $t_k$, we compute the matrix $U_k$ and the vector $V_k$, we solve the equation $U_k \alpha_k = V_k$ in order to obtain the vector of coefficients $\alpha_k$. It is possible to avoid making a regression at each date, by including the time in the regression. Discounted cash flows $\widetilde{P}_{k+1}$ are regressed against the state variable vector $X_k$ and against the time $t_k$. This means that the basis functions include the time $t_k$ as a variable. $f_{k,l}(X_k)$ is generalized to $f_l(X, t)$:
\begin{equation*}
C_k^{(j)} \simeq \widehat{C}_k\Big(X_k^{(j)}\Big) = \sum_{l=1}^p \alpha_l f_l(X_k^{(j)}, t_k) 
\end{equation*}
In this general case, we minimize the error function
\begin{equation*}
\Psi(\alpha) =  \sum_{k=1}^M \EE\Bigg[w_k(X_k) \bigg(C_k - \sum_{l=1}^p \alpha_l f_l(X_k,t_k) \bigg)^2\Bigg] \rlap{\ .} \end{equation*}
Similarly to what is explained is section \ref{sec:LeastSquare} we build a $p \times p$ matrix and a dimension $p$ vector
\begin{eqnarray}
U &=& \sum_{j=1}^N \sum_{k=1}^M w_k\Big(X_k^{(j)}\Big) f\Big(X_k^{(j)}\Big) f^\top\!\Big(X_k^{(j)}\Big)
\nonumber \\
V &=& \sum_{j=1}^N \sum_{k=1}^M w_k\Big(X_k^{(j)}\Big) f\Big(X_k^{(j)}\Big) \widetilde{P}_{k+1}^{(j)}
\rlap{\ .}
\label{eq:UV}
\end{eqnarray}
Then we solve the linear equation
$
U \alpha =  V
$
and get least squares coefficients
$
\alpha = U^{-1} V \rlap{\ .}
$

When the number of basis function is large, solving the linear system can be time-consuming if matrix $U$ is dense. However we can choose basis functions so that $U$ is block-diagonal. This is obtained if basis functions are divided in subsets with disjoint supports. To be more precise, let us assume we have $B$ blocks, labeled by $b$. We denote by $p_b$ the number of basis functions in block $b$, with $\displaystyle \sum_{b=1}^B p_b = p$. Inside block $b$, we denote basis functions by $f_{b,l}$ with $1 \leq l \leq p_b$. Functions which belong to two different blocks have disjoint support on $(X,t)$. Therefore, if $b \neq b'$ for all $X$ and $t$ we have $f_{b,l}(X,t) f_{b',l'}(X,t) = 0$. From the definition of matrix $U$ in equations \eqref{eq:UV} this means $U$ is block-diagonal. We denote by $f_b$ the vector of basis functions in block $b$, $U_b$ the diagonal blocks of matrix $U$, with a similar split of vector $V$ in $V_b$:
\begin{eqnarray*}
U_b &=& \sum_{j=1}^N \sum_{k=1}^M w_k\Big(X_k^{(j)}\Big) f_b\Big(X_k^{(j)}\Big) f_b^\top\!\Big(X_k^{(j)}\Big)
\nonumber \\
V_b &=& \sum_{j=1}^N \sum_{k=1}^M w_k\Big(X_k^{(j)}\Big) f_b\Big(X_k^{(j)}\Big) \widetilde{P}_{k+1}^{(j)}
\rlap{\ .}
\end{eqnarray*}

The classical date by date regression is the special case where a block corresponds to a given exercise date and where basis function are $f_{b,l}(X,t) = f_{b,l}(X) \mathbbm{1}_{t=t_b}$.

An other possibility, which requires fewer basis functions, is to partition the total set of exercise dates in $B$ groups of consecutive dates, with basis functions of a given block concentrated on the corresponding dates and null for other dates. If block $b$ corresponds to exercise dates $t_k$ with $k_{b-1} < k \leq k_b$, basis functions are taken of the form $f_{b,l}(X,t) = \tilde{f}_{b,l}(X,t) \mathbbm{1}_{t_{k_b-1} < t < t_{k_b}}$. As an example, if we have a set of $\tilde{p}$ basis functions $\tilde{f}_l(X)$ in the $X$ variable, we can construct a basis of functions with affine dependence on $t$ with 
\begin{eqnarray*}
f_{b,2l-1}(X,t) &=& \tilde{f_l}(X) \mathbbm{1}_{t_{k_b-1} < t < t_{k_b}}
\\
f_{b,2l}(X,t) &=& t \tilde{f_l}(X)\mathbbm{1}_{t_{k_b-1} < t < t_{k_b}}
\end{eqnarray*}
Thanks to that, the coefficients $\alpha_k$ won't be computed at all dates. We will have only a matrix $U_b$ and the vector $V_b$ for a set of exercise times $[t_{k_{b-1}+1}, \ldots, t_{k_b}]$. In addition, this can reduce the statistical error on the exercise boundary: for a given number of paths there are more contributions in $U$ and $V$.

\section{Numerical results}
\label{sec:results}

We consider the example of an American put on an asset $S_t$. Assume that the stock price follows the Black-Scholes dynamic and that there is no arbitrage opportunity. The risk-neutral process of the stock price is the following:
\begin{equation*}
\ud S_t = r S_t \ud t + \sigma S_t \ud W_t \rlap{\ .}
\end{equation*}
The risk-less interest rate $r$ and the volatility $\sigma$ are assumed to be constant. There are no dividends. We denote by $K$ the strike price and by $T$ the maturity of the option. 

We use the same example as in \cite{longstaff2001valuing}. We price an American put option on a share with strike price \$40. The annual interest rate is 6\%, the underlying stock price is \$36, the volatility $\sigma$ is 20\% and the maturity 1 year. We consider that the option can be exercised 50 dates per year until its maturity.

We generate 100,000 paths. In the parallel algorithm, we use 100 iterations, independently of the total number of paths. We choose 5 groups of 10 dates. The basis functions chosen for the regression are : $1$, $S$, $S^2$, $t$, $tS$ and $tS^2$. 
We have weights on $U$ and $V$, $w_{UV}$ with $\lambda=2$ and $\mu=2$. We use weights on prices $w_i$ with $\nu=0.99$. We also have the weight depending on the path $y_k$.

We compare results with a reference value of \$4.486 given by a finite difference method.  We use an implicit scheme with 40,000 time steps and 1,000 steps for the stock price.

\subsection{Convergence of the algorithm}
We have implemented the parallel algorithm and we have compared it with the finite difference method. We have tested the impact of the number of iterations and the number of dates per block. The finite difference American is the result of a the discretization of the Black-Scholes equation:
\begin{equation*}
\partial_t P + \frac{1}{2}\sigma^2 S^2 \partial_{S^2}P + r S \partial_S P - r P = 0
\end{equation*}
with the terminal condition $P(T,S) = \max(0, S-K)$.

\subsubsection{Number of iterations}
Our example is tested on a quad-core CPU. We parallelize the algorithm on four threads. In each thread, the paths are generated and the matrices $U_b$ and vectors $V_b$ are computed for each date $k$. For each thread, we only need to keep $U_b$, $V_b$ for all $b$ and the sum of the present value. When computation is finished in all threads, the results are aggregated. When we have the global $U_b$ and $V_b$ which are the sum of all the matrices $U_b$ and vectors $V_b$ of each thread, the coefficients $\alpha_b$ of the regression are computed by solving $U_b \alpha_b = V_b$. This step is also done in parallel by solving this equation for a block of dates $b$ in each thread. When the coefficients are computed, we use them in the following iteration for the computation of $U_k$ and $V_k$ and also the option price. In the first iteration we do not have the $\alpha_k$ needed. We make the decision to keep the option until its maturity. We could also use coefficients from the previous day computation. 

\begin{figure}[!h]
\begin{center}		
  \includegraphics[width=1\textwidth]{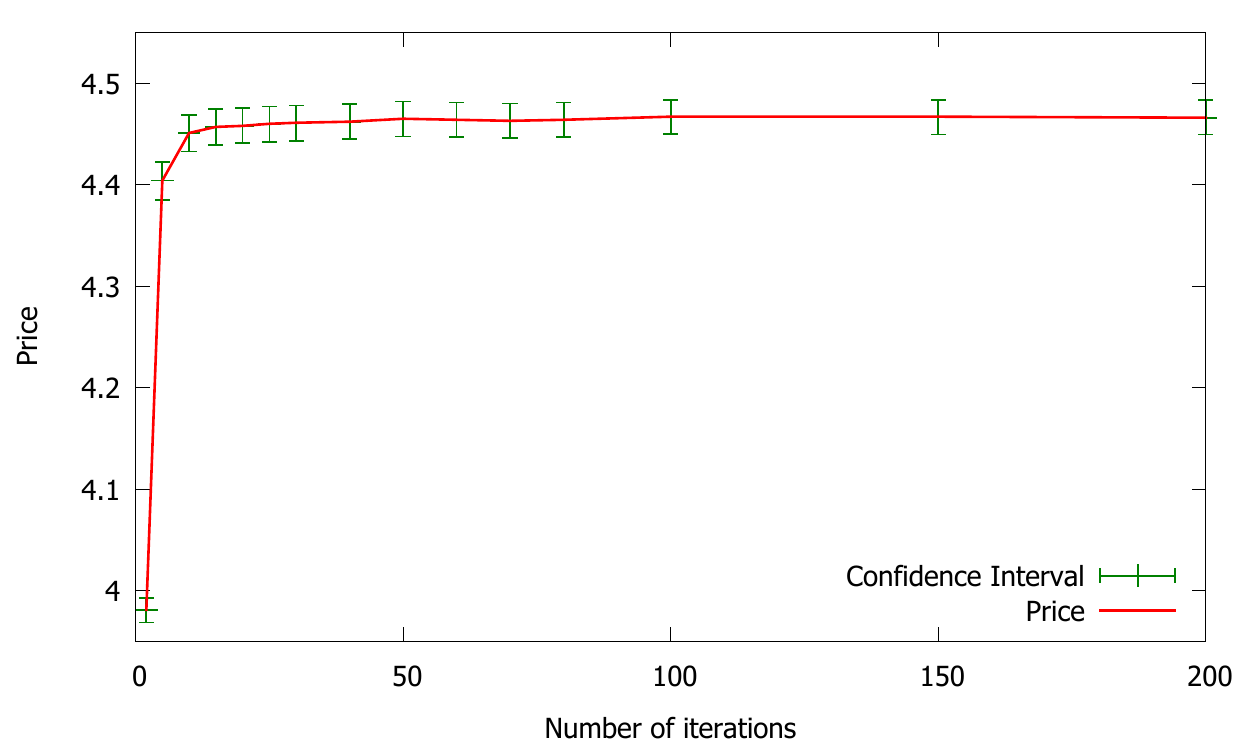}
	\caption{The impact of the number of iterations for a given number of paths (100,000) on the price.}
	\label{plot:it}
\end{center}
\end{figure}

Figure \ref{plot:it} 
shows the impact of the number of iterations on the final price. In this figure, the total number of paths generated remains the same, only the number of paths per iteration changes.

During the Monte Carlo pricing we compute the (weighted) variance of prices $\mathcal{V} = \frac{1}{q_N} \sum_{i=1}^n \widetilde{w}_i \sum_{j=n_{i-1}+1}^{n_i} {P_1^{(j)}}^2 - P^2$ with $q_N = \sum_{i=1}^n \widetilde{w}_i (n_i-n_{i-1})$. Using also $q_N^{(2)} = \sum_{i=1}^n \widetilde{w}_i^2 (n_i-n_{i-1})$ we get the standard error estimate $\varepsilon = \sqrt{\mathcal{V} \frac{q_N^{(2)}}{q_N^2}}$. We plot the statistical 95\% confidence interval, which corresponds to $\pm 1.96 \varepsilon$. Note that it takes into account statistical error only and not systematic error.

\begin{figure}[!hp]
\begin{center}		
  \includegraphics[width=1\textwidth]{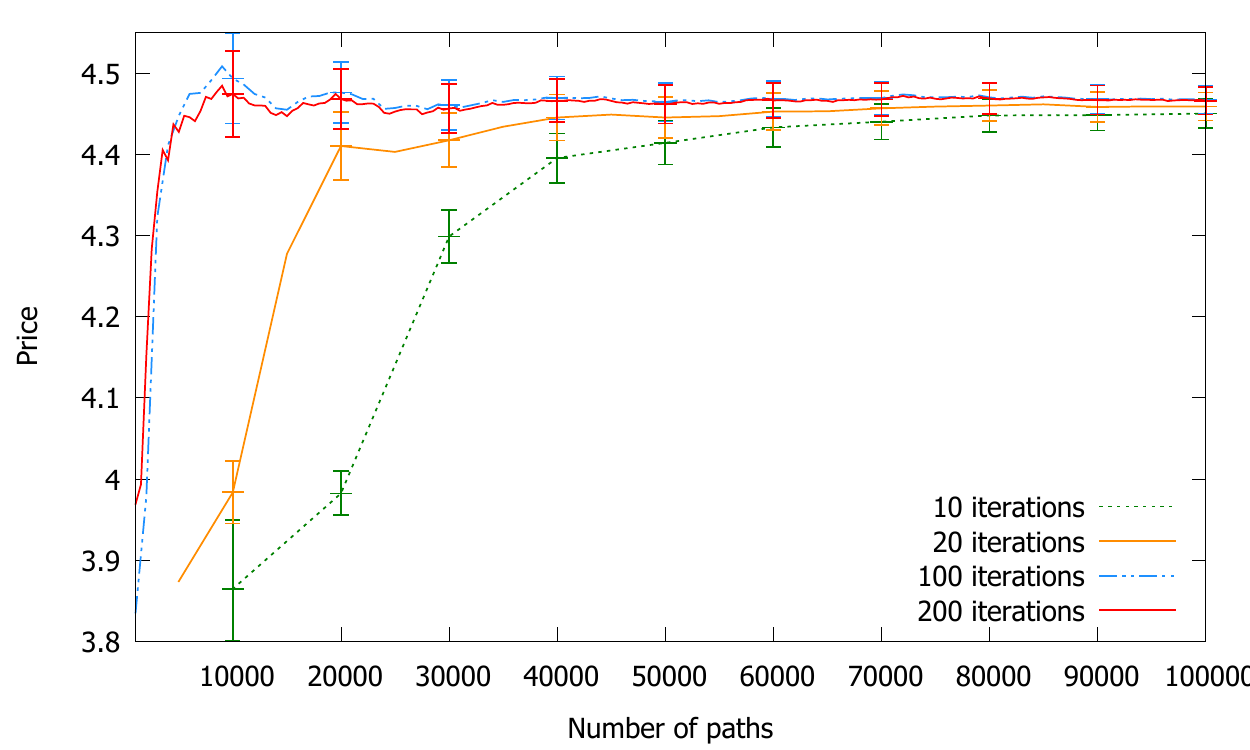}
	\caption{The impact of the number of iterations on the American put price.}
	\label{plot:PriceIterations}
\end{center}
\end{figure}

\begin{figure}[!hp]
\begin{center}
  \includegraphics[width=1\textwidth]{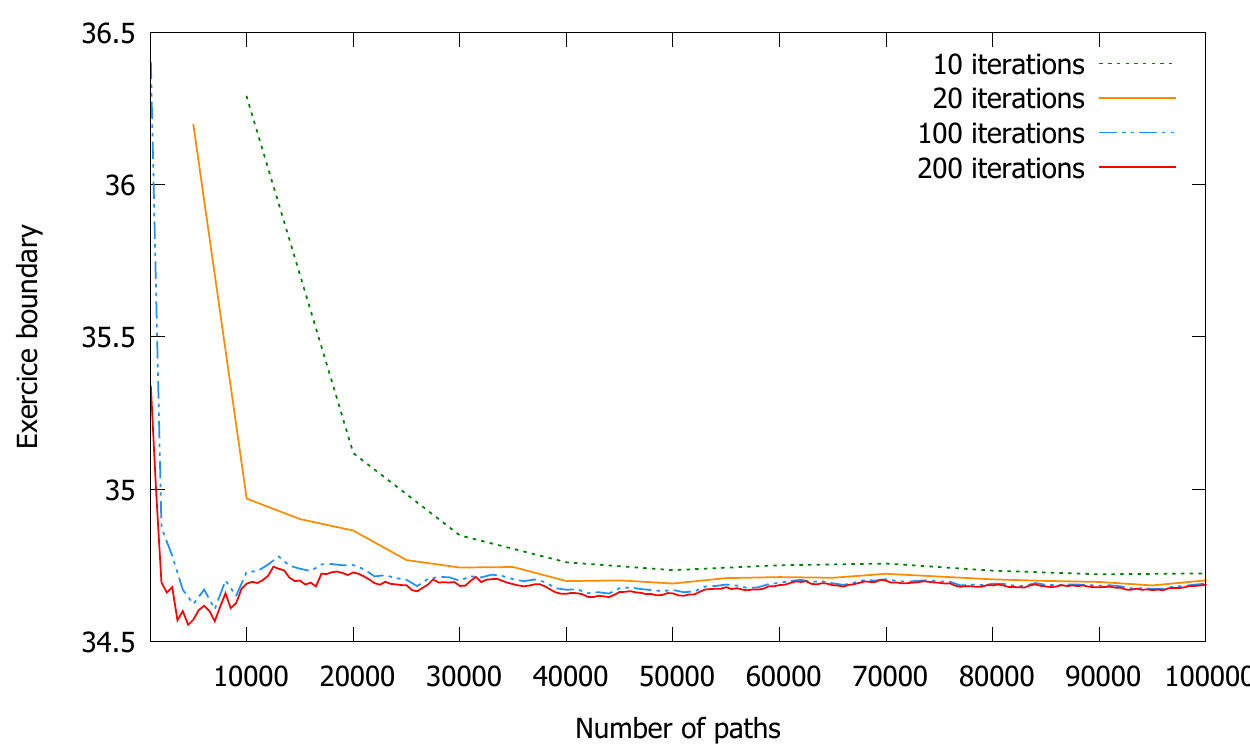}
	\caption{The impact of the number of iterations on the American put early exercise boundary at the mid-maturity date.}
	\label{plot:tBoundaryIterations}
\end{center}
\end{figure}

The price converges closer to the real price \$4.486 when the number of iterations increases. We notice that for 100,000 paths, 100 iterations are sufficient to converge. Going further, figure \ref{plot:PriceIterations} presents the price convergence for different numbers of iterations $[10, 20, 100, 200]$. Similarly, figure \ref{plot:tBoundaryIterations}
shows the convergence of the early-exercise boundary at the mid-maturity date. The convergence is faster for a larger number of iterations. However the difference between 100 and 200 iterations is not significant. In these two cases, a good price estimate is obtained after 10,000 paths. In addition, we notice that for 100,000 paths, the price obtained with only 10 iterations is different from the price with 200 iterations by less than two standard errors.

\subsubsection{Weights for $U,V$ and price} 
As the algorithm is iterative, the values of the regressions coefficients and of the price are not correct for the first iterations. We have added the rescaling factor $w_{UV}^{(i)}$ from equation \eqref{eq:wUV} with $\lambda=2$ and $\mu=2$.
Each $U_k$ and $V_k$ from previous iteration are multiplied by $w_{UV}^{(i)}$.

\begin{figure}[!hp]
\begin{center}		
  \includegraphics[width=1\textwidth]{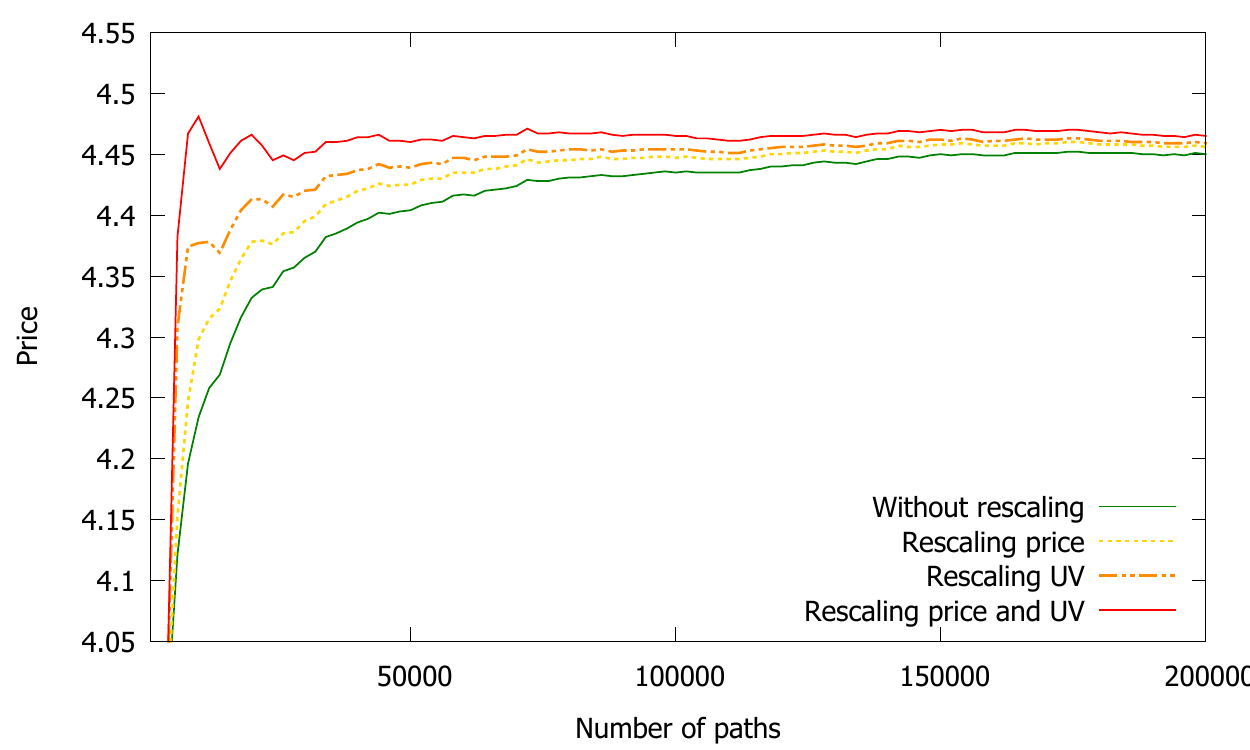}
	\caption{The impact of weighting the price or $U,V$ for each iteration on the American put price.}
	\label{plot:weightPrice}
\end{center}
\end{figure}

\begin{figure}[!htp]
\begin{center}

  \includegraphics[width=1\textwidth]{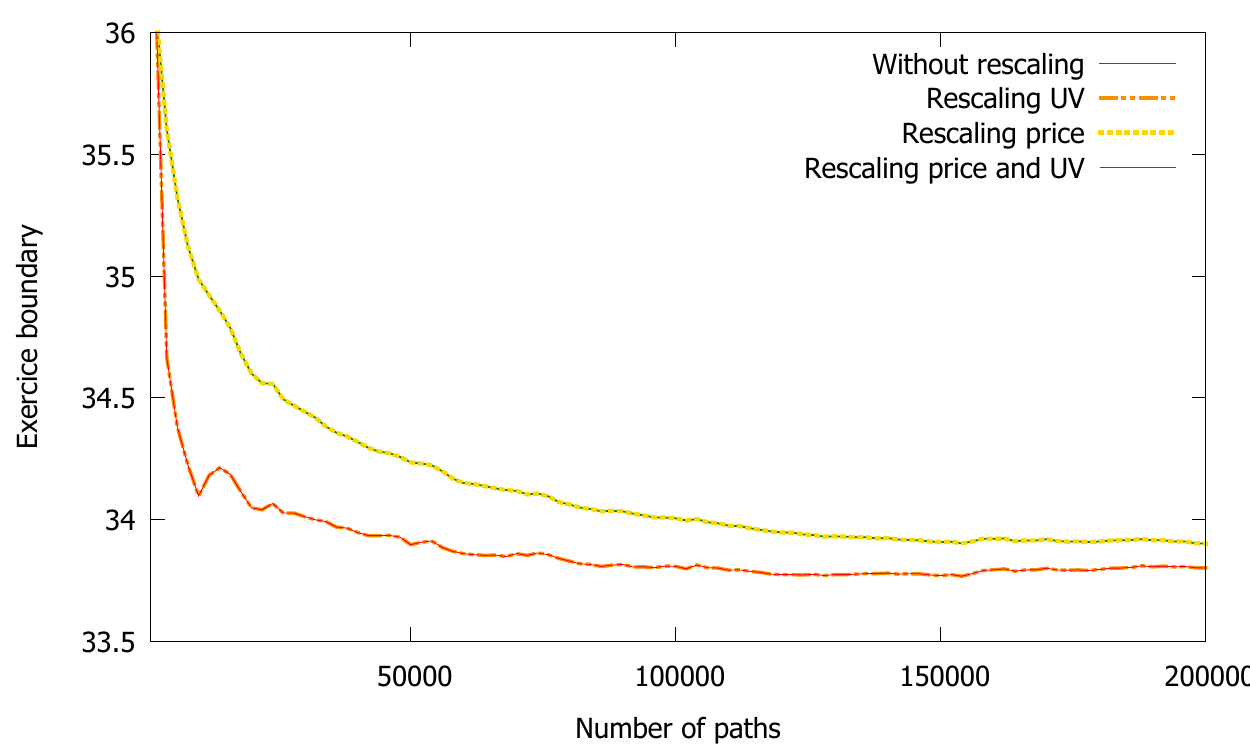}
	\caption{The impact of weighting the price or $U,V$ for each iteration on the American put early exercise boundary at mid-maturity.}
	\label{plot:weightBoundary}
\end{center}	
\end{figure}

In the same way, we add a  weight on the price that depends on the number of iteration $\widetilde{w}_i$ from equation \eqref{eq:wP} with $\nu=0.99$.
At each iteration $i$, we multiply the sum of present values of the paths in the iteration by $\widetilde{w}_i$ before adding to the sum of present values of the previous iterations. 
In figure \ref{plot:weightPrice} we show the impact of the various weights on the price. The price converges faster if we add weights in both $U$, $V$ and in the price. We also plot an early exercise boundary in figure \ref{plot:weightBoundary}.

It corresponds to the boundary at the mid-maturity date.
One can see that the weight of $U$ and $V$,  $w_i$ has an impact on the boundary but not the weight of the price $\widetilde{w}_i$. This is due to the fact that $w_i$ has an impact on the coefficients $\alpha_b$ of the regression which are used in the computation of the exercise boundary. On the opposite, the weight on the price $\widetilde{w}_i$ does not have an impact on the boundary, as the rescaling is done on the price alone, after the computation of the coefficients and exercise boundaries. 

\subsubsection{Size of date groups}

In the algorithm of Longstaff-Schwartz, a regression is made at each date $t_k$. We choose as basis functions $1$, $S$ and $S^2$.  The continuation value is estimated as
 $$E[\tilde{P}(S_{t+1}) | S_t] \simeq \alpha + \beta S_{t} + \gamma S_{t}^2 \rlap{\ .}$$ The coefficients are computed at each time $t_k$ in  $[t_1, ... , t_M]$. We include the time in the regression variables and we add three more basis functions: $t$, $tS$ and $tS^2$: 
 $$E[\tilde{P}(S_{t+1}) | S_t , t] = \alpha +\beta S_{t} +\gamma S_{t}^2 + \delta t  + \varepsilon t S_t + \zeta t S_t^2 \rlap{\ .}$$

\begin{figure}[!htp]
\begin{center}		
  	\includegraphics[width=1\textwidth]{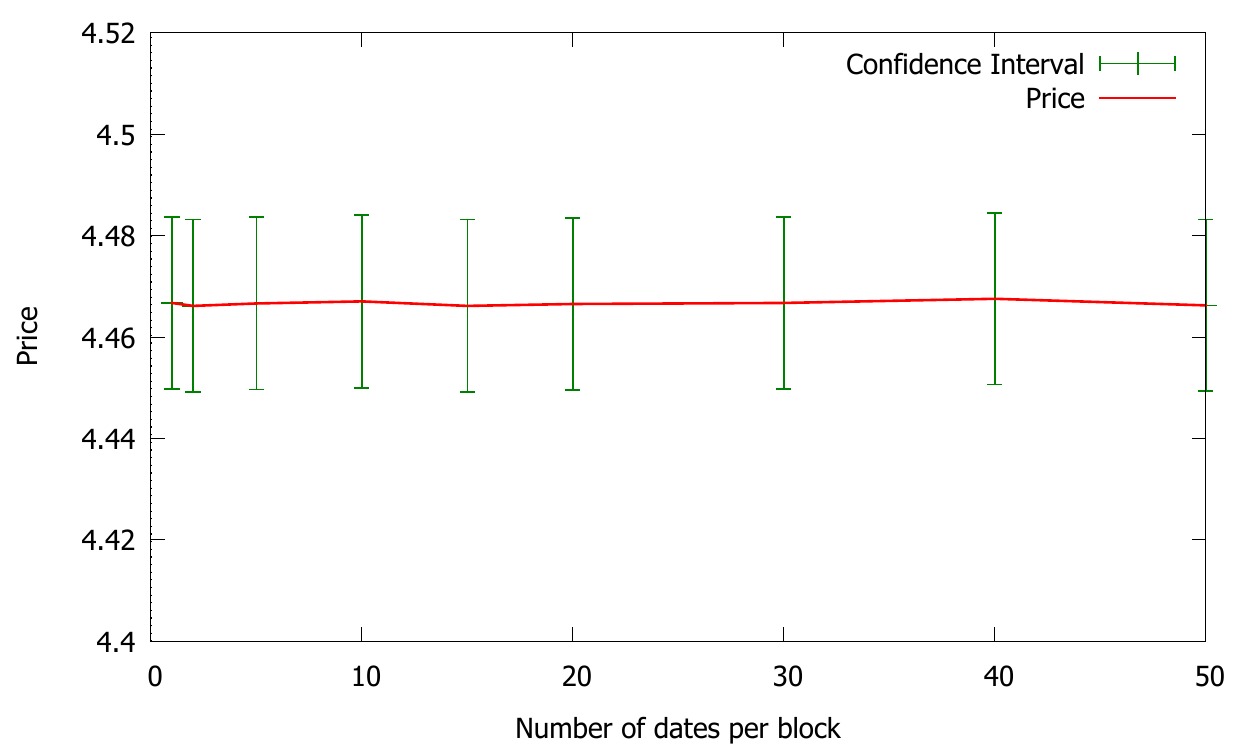}
	\caption{The impact of the size of the dates groups on the American put price.}
	\label{plot:dates}
\end{center}
\end{figure}

We make groups of $D$ dates $[t_{bD-D+1}, ... , t_{bD}]$. The resolution of the equation  $U_b \alpha_b = V_b$ is made only once per group of dates. With the coefficients computed for one group $b$, we can estimate the discounted value $\tilde{P}$ for all dates within the group $[t_{bD-D+1}, ... , t_{bD}]$.
We have tested for several sizes of dates groups. As figure \ref{plot:dates}
shows, the number of dates per group does not have an important impact on the price. 
In the graph, we also have the case of one date per group, which means that we are in the first case with three basis functions. The price estimate is very similar in both cases.
With more dates per group, the total number of groups is reduced and thus also the number of linear systems to inverse. Therefore, using groups of dates may save some computation time and reduce the quantity of data to transfer without deteriorating the precision of the price.

\subsection{Comparison with Longstaff Schwartz}
\begin{figure}[!hp]
\begin{center}		
  \includegraphics[width=1\textwidth]{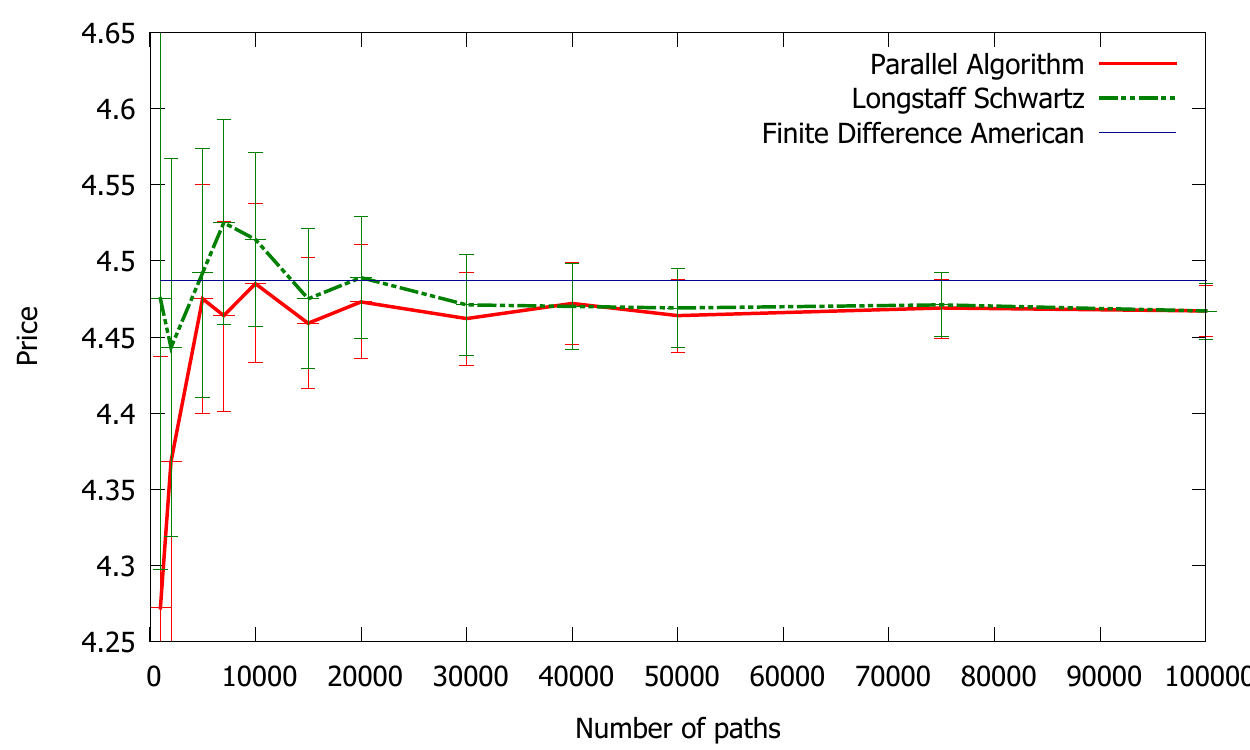}
	\caption{Convergence of Longstaff Schwartz vs Parallel Algorithm.}
	\label{plot:cv}
\end{center}
\end{figure}

In this section we compare our parallel algorithm with the Longstaff-Schwartz algorithm, using the same example and parameters. We show the price for different numbers of paths in figure \ref{plot:cv}.

Both algorithms converge to the same price which is below the \$4.486 price obtained with finite difference method by 1.9{\textcent} (0.4\% relative error). As we explained in section \ref{eq:convergence} this is due to the approximation of the continuation value which makes the exercise slightly sub-optimal.

\begin{figure}[!htp]
\begin{center}		
      \includegraphics[width=1\textwidth]{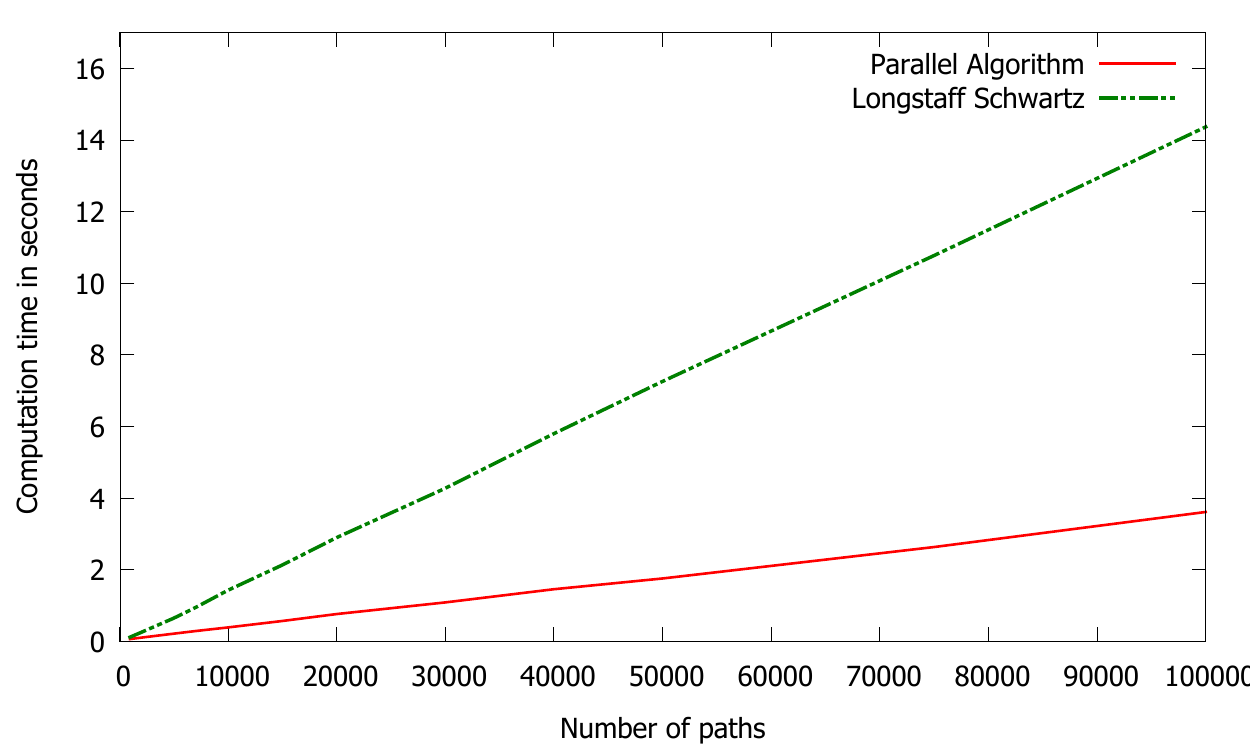}
	\caption{Computation time of Longstaff-Schwartz vs Parallel Algorithm with 4 cores.}
	\label{plot:cvspeed}
\end{center}
\end{figure}

\begin{figure}[!htp]
\begin{center}	
      \includegraphics[width=1\textwidth]{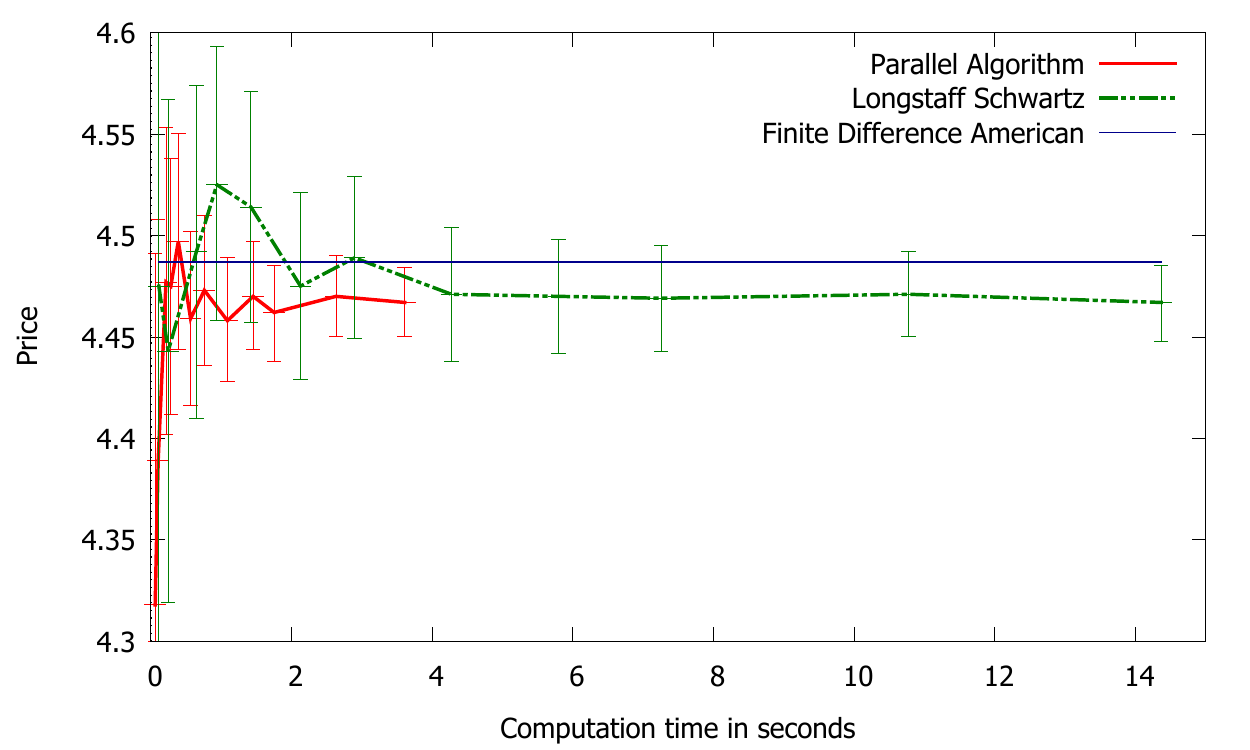}
	\caption{The convergence of the price with respect to the computing time.}
	\label{plot:cvspeedinTime}
\end{center}
\end{figure}

What is remarkable and innovative is that the parallel algorithm is using all available threads (in our example, four) during the whole computation. The Longstaff-Schwartz algorithm uses only one thread. Thus for 100,000 paths the Longstaff-Schwartz needs 14.37 seconds while the parallel algorithm takes only 3.6 seconds, as shown in figure \ref{plot:cvspeed}.
One observes a good scaling property. Even if one parallelizes the path generation step in the LSM, we still have an important improvement with our algorithm\footnote{In our example, path generation takes 8.42 seconds over a total of 14.37 seconds in LSM. Parallelizing this step would give a total computation time of at least 8.05 seconds versus 3.6 seconds with our algorithm. This is without taking in consideration the memory issues and the data transfer cost.}.
Figure \ref{plot:cvspeedinTime} plots the price estimate against the computing time for both algorithm in our quad-core example.

In table \ref{tab:put} we compare the price of American put options on a share using the Longstaff-Schwartz algorithm, the parallel algorithm and the finite difference method.
\begin{sidewaystable}[p]
\begin{center}
\small\addtolength{\tabcolsep}{-3pt}
\begin{tabular}{l*{15}{c}r}
	&		&		&	Finite		&	Least		&					&	Parallel		&		Parallel		&	Closed		&	Early		&	Early	&	Early	&	Difference	&	Difference	&	Difference	\\
	&		&		&	Difference	&	Squares		&		LS			&	Algorithm		&		Algorithm		&	Formula		&	Exercise		&	Exercise	&	Exercise	&	PDE and		&	PDE and		&	LS and	\\
S	& 	$~~~~\sigma~~~~$	&	T	&	American	&	Simulation	&		(s.e)		&	Simulation	&		(s.e)		&	European	&	PDE			&	 LS		&	Parallel	&	LS			&	Parallel		&	Parallel	\\\hline																																						\\
36	&	0.2	&	1	&	4.486	&	4.467	&	(.009)	&	4.467	&	(.009)	&	3.844	&	.642	&	.622	&	.623	&	.019	&	.019	&	.000	\\
36	&	0.2	&	2	&	4.847	&	4.833	&	(.011)	&	4.838	&	(.011)	&	3.763	&	1.084	&	1.070	&	1.075	&	.014	&	.009	&	-.005	\\
36	&	0.4	&	1	&	7.109	&	7.087	&	(.019)	&	7.100	&	(.019)	&	6.711	&	.398	&	.376	&	.389	&	.022	&	.009	&	-.013	\\
36	&	0.4	&	2	&	8.513	&	8.512	&	(.022)	&	8.521	&	(.022)	&	7.700	&	.813	&	.812	&	.821	&	.001	&	-.008	&	-.009	\\
																																									\\
38	&	0.2	&	1	&	3.257	&	3.237	&	(.009)	&	3.247	&	(.009)	&	2.852	&	.405	&	.385	&	.395	&	.020	&	.009	&	-.011	\\
38	&	0.2	&	2	&	3.750	&	3.738	&	(.011)	&	3.753	&	(.011)	&	2.991	&	.760	&	.747	&	.762	&	.012	&	-.003	&	-.015	\\
38	&	0.4	&	1	&	6.155	&	6.145	&	(.019)	&	6.151	&	(.018)	&	5.834	&	.320	&	.310	&	.317	&	.010	&	.004	&	-.007	\\
38	&	0.4	&	2	&	7.674	&	7.663	&	(.022)	&	7.678	&	(.022)	&	6.979	&	.696	&	.684	&	.699	&	.011	&	-.003	&	-.015	\\
																																									\\
40	&	0.2	&	1	&	2.319	&	2.305	&	(.009)	&	2.306	&	(.008)	&	2.066	&	.253	&	.239	&	.240	&	.014	&	.013	&	-.001	\\
40	&	0.2	&	2	&	2.889	&	2.870	&	(.011)	&	2.878	&	(.010)	&	2.356	&	.533	&	.515	&	.523	&	.019	&	.011	&	-.008	\\
40	&	0.4	&	1	&	5.319	&	5.306	&	(.018)	&	5.310	&	(.018)	&	5.060	&	.259	&	.247	&	.251	&	.012	&	.008	&	-.004	\\
40	&	0.4	&	2	&	6.923	&	6.918	&	(.022)	&	6.924	&	(.021)	&	6.326	&	.597	&	.592	&	.598	&	.005	&	-.001	&	-.006	\\
																																									\\
42	&	0.2	&	1	&	1.621	&	1.615	&	(.008)	&	1.613	&	(.007)	&	1.465	&	.157	&	.151	&	.149	&	.006	&	.008	&	.002	\\
42	&	0.2	&	2	&	2.216	&	2.194	&	(.010)	&	2.204	&	(.009)	&	1.841	&	.375	&	.353	&	.362	&	.022	&	.012	&	-.010	\\
42	&	0.4	&	1	&	4.589	&	4.591	&	(.017)	&	4.584	&	(.017)	&	4.379	&	.210	&	.212	&	.205	&	-.003	&	.005	&	.007	\\
42	&	0.4	&	2	&	6.250	&	6.241	&	(.021)	&	6.247	&	(.021)	&	5.736	&	.514	&	.506	&	.511	&	.009	&	.003	&	-.006	\\
																																									\\
44	&	0.2	&	1	&	1.113	&	1.112	&	(.007)	&	1.109	&	(.006)	&	1.017	&	.096	&	.095	&	.092	&	.001	&	.004	&	.003	\\
44	&	0.2	&	2	&	1.693	&	1.680	&	(.009)	&	1.686	&	(.009)	&	1.429	&	.264	&	.251	&	.257	&	.013	&	.007	&	-.006	\\
44	&	0.4	&	1	&	3.953	&	3.959	&	(.016)	&	3.952	&	(.016)	&	3.783	&	.171	&	.177	&	.169	&	-.006	&	.001	&	.007	\\
44	&	0.4	&	2	&	5.647	&	5.651	&	(.021)	&	5.651	&	(.020)	&	5.202	&	.445	&	.449	&	.449	&	-.004	&	-.004	&	.000	\\
\hline	
\end{tabular}
\end{center}
\caption{Comparison of the American put prices.}
\label{tab:put}	
\end{sidewaystable}
We use the same parameters as in the previous example. We compute the price for different values of the underlying spot price $S=36, 38, 40, 42, 44$, of the volatility $\sigma=20\%, 40\%$ and of the maturity $T=1, 2$. In this table, we also present the standard error (s.e) for each algorithm, the price of a European put option and the early exercise value which is the difference between the American and the European price.

The differences between the finite difference and the LSM algorithm are very small. The 20 differences are less or equal to 2.2{\textcent}, among which 9 values are less or equal to 1{\textcent}. The standard error for the simulated value ranges from 0.7{\textcent} to 2.2{\textcent}. The differences of the finite difference and the parallel algorithm are even smaller. The 20 differences are less or equal to 1.9{\textcent}, among which 16 values are less or equal to 1{\textcent}. The standard errors are similar to the LSM standard errors, 0.6{\textcent} to 2.2{\textcent}. All differences between the LSM and the parallel algorithm are smaller than one standard error. The differences with the finite difference are both positive and negative for both algorithms.

\subsection{Improved exercise decision in the first iteration}

At each iteration, the exercise strategy is determined by the coefficients coming from the previous iterations. In the first iteration, the coefficients are not available. Therefore, for the first iteration, the choice made in our previous examples was to exercise the option at the maturity.

\begin{figure}[!hp]
\begin{center}		
  \includegraphics[width=1\textwidth]{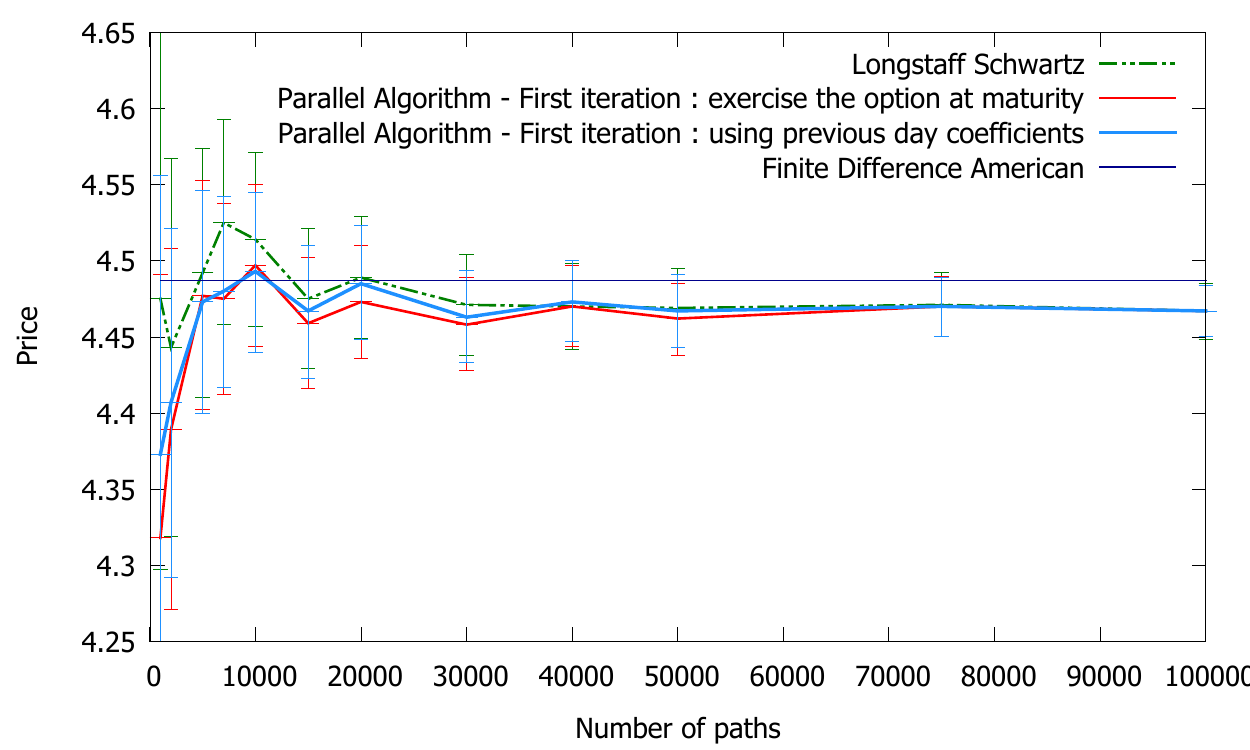}
	\caption{Convergence of Longstaff-Schwartz vs Parallel Algorithm using the coefficients of the previous day for the first iteration.}
	\label{plot:cvld}
\end{center}
\end{figure}

\begin{figure}[!hp]
\begin{center}		
  \includegraphics[width=1\textwidth]{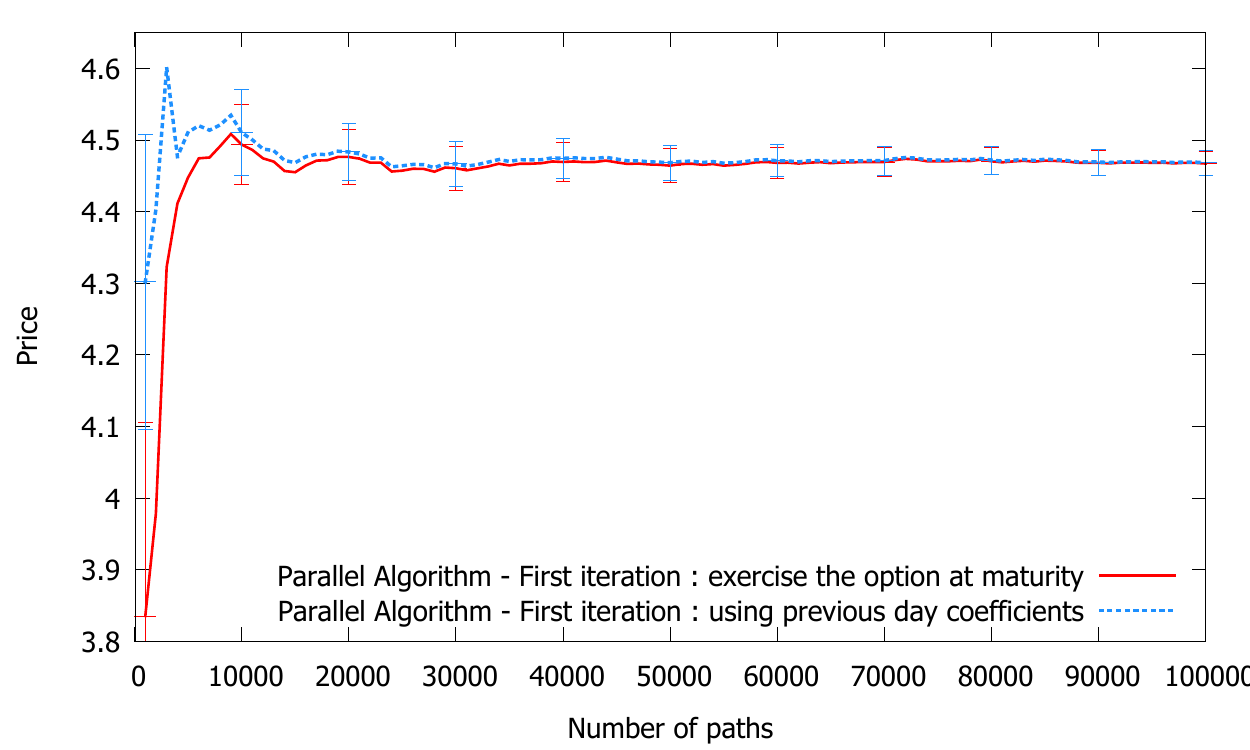}
	\caption{The evolution of the American put price at each iteration  for both exercise strategies in the first iteration.}
	\label{plot:PriceLastDay}
\end{center}
\end{figure}

\begin{figure}[!hp]
\begin{center}	
  \includegraphics[width=1\textwidth]{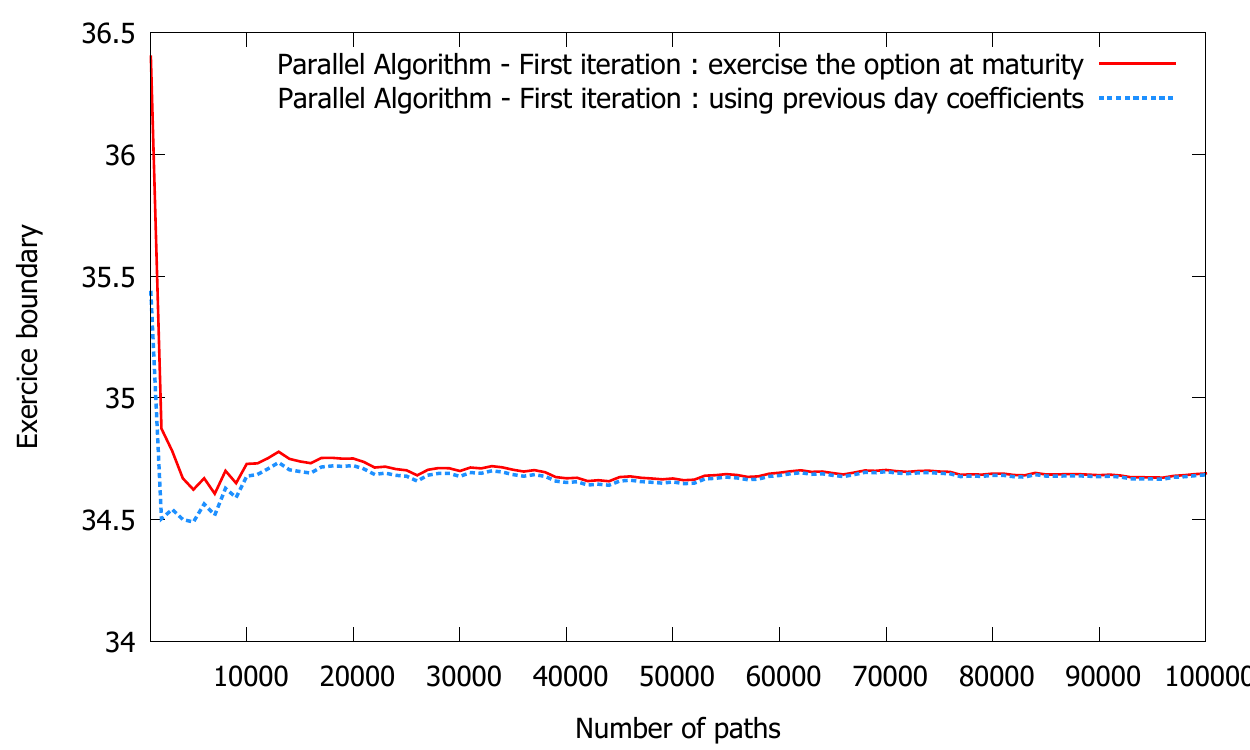}
	\caption{The evolution of the American put early exercise boundary at the mid-maturity date at each iteration for both exercise strategies in the first iteration.}
	\label{plot:tBoundaryLastDay}
\end{center}
\end{figure}

Another solution is to use the coefficients of the previous computation, which is usually made the previous day. We illustrate this case in the figures \ref{plot:cvld}, \ref{plot:PriceLastDay} and \ref{plot:tBoundaryLastDay}.

In this example for the first iteration only we use the coefficients and therefore the exercise boundaries computed in a previous computation, with different market parameters. The interest rate is $5.5 \%$, the volatility $\sigma$ is $22\%$ and the spot value is \$34. 

Figure \ref{plot:cvld} shows the convergence of the put price.
We launch several times the pricing with increasing number of paths. We observe that using previous day coefficients for the first iteration improves the convergence of the algorithm.

Going further, figures \ref{plot:PriceLastDay} and  \ref{plot:tBoundaryLastDay} show the evolution of the price and of the mid-maturity early exercise boundary during the computation of one pricing.
It displays the price and boundary values after each iteration. The price using the previous day coefficients for the first iteration is higher and closer to the correct price in the first iteration. Without exercise until maturity in the first iteration, we notice that the price of the American put has the value of an European put of \$3.844 in the first iteration. After a few iterations it converges to the American price.

In figure \ref{plot:tBoundaryLastDay} we see that the exercise boundary is higher for the first iteration than the following ones in both cases.
This is explained because the exercise is not optimal in the first iteration, and therefore the continuation values are underestimated. As we consider a put option, this means that the boundaries are estimated higher than their real value. This phenomenon is reduced with coefficients from the previous day in the first iteration due to more optimal exercise.

As a summary, we propose two alternatives methods for the first iteration: starting with an European option or using the previous day coefficients. This last method improves the convergence of the algorithm as we use a starting point closer to the real exercise strategy.

\section{Conclusion}
\label{sec:conclusion}

This article introduces a new algorithm for pricing American options or callable structured products by simulations, using least squares regression. It can also be used to compute counterparty credit risk like CVA or PFE. This algorithm is intuitive, easy to implement and attractively scalable as it can be fully parallelized. The computing time is almost divided by the number of calculators. There is no need to store the paths and the computation can be done forwards. This allows to price derivatives where exercise decisions depend non-trivially on previous decisions.
\begin{appendices}
\section{Continuation value}
\label{app:continuation}
\begin{proof}
Expanding the square and using linearity of the expected value, we can rewrite the error function $\Psi_k$ as
\begin{multline*}
\Psi(\alpha_{k}) = \EE\Big[w_k(X_k) \EE\big(\widetilde{P}_{k+1} \big| X_k\big)^2 \Big] - 2 \, \EE\Bigg[w_k(X_k) \EE\big(\widetilde{P}_{k+1} \big| X_k\big) \sum_{l=1}^p \alpha_{k,l} f_{k,l}(X_k)\Bigg] 
\\
+ \EE\Bigg[w_k(X_k) \bigg(\sum_{l=1}^p \alpha_{k,l} f_{k,l}(X_k)\bigg)^2\Bigg] \ .
\end{multline*}
In the righthand side, there are three terms in the expected value. The first one is quadratic but does not depend on $\alpha_{k,l}$: it is a constant which is not relevant in the minimization problem. We can replace it by the other constant term $\EE\Big[\widetilde{P}_{k+1}^2  \Big]$: the minimum will be shifted but the coefficients $\alpha_{k,l}$ which minimize the function will be the same. The second term can be rewritten as
\begin{multline*}
\EE\Bigg[w_k(X_k) \EE\big(\widetilde{P}_{k+1} \big| X_k\big) \sum_{l=1}^p \alpha_{k,l} f_{k,l}(X_k)\Bigg]
\\
=
\EE\Bigg[\EE\bigg(w_k(X_k)  \widetilde{P}_{k+1}  \sum_{l=1}^p \alpha_{k,l} f_{k,l}(X_k)\bigg| X_k\bigg) \Bigg]
\\
=
\EE\Bigg[w_k(X_k) \widetilde{P}_{k+1}  \sum_{l=1}^p \alpha_{k,l} f_{k,l}(X_k) \Bigg] \rlap{\ .}
\end{multline*}
Keeping the last term as it is and refactoring the three terms, we find that minimizing $\Psi_k$ is equivalent to minimizing $\Phi_k$.
\end{proof}

\section{Convergence}
\label{app:convergence}
\begin{proof}
Let us assume there are $m$ paths per iteration and $n$ iterations. We denote collectively by $\alpha_i$ the vector of regression coefficients computed in the iteration $i$. 
We denote by $u_i$ and $v_i$ the average contribution of paths from iteration $i$ to matrices $U$ and $V$ of the least squares regression \eqref{eq:UV-LS}. $u_i$ and $v_i$ depend on the coefficients computed from the previous iteration $\alpha_{i-1}$ and on the random variables used to compute the paths in iteration $i$, that we denote collectively by $\varepsilon_i$. In order to simplify the notation we denote by $\phi$ the functions $u$ and $v$ simultaneously. The contribution $\phi_i$ is the average of $\phi$ on the $m$ paths of the $i^{th}$ iteration.
\begin{equation}
\phi_i = \phi(\alpha_{i-1},\varepsilon_i) = \frac{1}{m}\sum_{j=1}^m \phi(\alpha_{i-1},\varepsilon_i^j) 
\end{equation}
 We decompose the matrix-valued function $u$ and the vector-valued function $v$ as the sum of their expected value $\bar \phi(\alpha) = \EE\big[ \phi(\alpha,\varepsilon) \big]$ and the stochastic part $\hat \phi(\alpha,\varepsilon) = \phi(\alpha, \varepsilon) - \bar \phi(\alpha) $ with null expected value :
\begin{equation}
\phi(\alpha, \varepsilon) = \bar \phi(\alpha) + \hat \phi(\alpha,\varepsilon)
\label{f_expsto}
\end{equation}
Let us consider the function
$$
\bar\alpha(\alpha) = \bar u(\alpha)^{-1} \bar v(\alpha) \rlap{\ .}
$$
We assume that the $\alpha \mapsto \bar\alpha(\alpha)$ is contractant:
$$
\forall \alpha, \alpha' \quad \| \bar\alpha(\alpha) - \bar\alpha(\alpha') \| \leq q \| \alpha - \alpha' \|
$$
with $q <1$.
From Banach fixed point theorem, it therefore admits a fixed point. We also denote it by $\bar\alpha$: 
\begin{equation}
\bar \alpha = \bar\alpha(\bar \alpha) = \bar u(\bar \alpha)^{-1} \bar v(\bar \alpha) \rlap{\ .}
\label{eq:fixedpoint}
\end{equation}
When the Longstaff-Schwartz algorithm can be used, it would correspond to the regression coefficients obtained with this algorithm in the limit of an infinite number of paths. Defining $\Delta \alpha = \alpha - \bar \alpha$, we write the Taylor expansion of the expected value $\bar \phi$ and of the stochastic part $\hat \phi$ around $\bar \alpha$.
\begin{eqnarray*}
\bar \phi(\alpha) &=& \bar \phi(\bar \alpha) + \frac{\partial \bar \phi(\alpha)}{\partial \alpha}\Big\vert_{\alpha=\bar\alpha} \Delta \alpha + O(\Delta \alpha^2) \\
\hat \phi(\alpha, \varepsilon) &=& \hat \phi(\bar \alpha, \varepsilon) + O_\phi(\Delta \alpha, \varepsilon) 
\end{eqnarray*} 
In order to simplify, let us call $\hat \phi(\varepsilon)$ the function $\hat \phi(\bar \alpha, \varepsilon)$. The decomposition of $\phi_i = \phi(\alpha_{i-1},\varepsilon_i)$ in \eqref{f_expsto} becomes:
\begin{equation}
\phi_i = \bar \phi(\bar\alpha) + \Delta \phi_i
\label{eq:uv-exp}
\end{equation}
with
\begin{eqnarray}
\Delta \phi_i &=& \frac{\partial \bar \phi(\alpha)}{\partial \alpha}\Big\vert_{\alpha=\bar\alpha} \Delta \alpha_{i-1} + \hat \phi(\varepsilon_i) + O(\Delta \alpha_{i-1}^2) + O_\phi(\Delta \alpha_{i-1}, \varepsilon_i) 
\label{eq:duv-order1}
\end{eqnarray}
We will focus only on the dominant terms and will not take in consideration the last two negligible elements $O(\Delta \alpha_{i-1}^2)$ and $O_\phi(\Delta \alpha_{i-1}, \varepsilon_i) $. 
Let us consider $\Phi_n$ the weighted average of $\phi_i$ up to the iteration $n$ with weights $w_i$, $\Phi_n = \frac{1}{z_n} \sum_{i=1}^n w_i \phi_i$
with $z_n = \sum_{i=1}^n w_i$. $\Phi_n$ is a notation for $U_n$ and $V_n$. Summing over expressions \eqref{eq:uv-exp} $\Phi_n$ reads
\begin{equation*}
\Phi_n = \bar \phi(\bar \alpha) + \Delta \Phi_n
\end{equation*}
with $\Delta \Phi_n = \frac{1}{z_n} \sum_{i=1}^n w_i \Delta \phi_i$.
Isolating the contribution from the latest iteration, this can be rewritten as a recursion:
\begin{equation}
\Delta \Phi_n = \frac{z_{n-1} \Delta \Phi_{n-1} + w_n \Delta \phi_n}{z_n} 
\label{eq:req-uv}
\end{equation}
After iteration $n$, the regression coefficients are computed as $\alpha_n = U_n^{-1} V_n$. Expanding around $\bar \alpha$ we have
\begin{eqnarray*}
\alpha_n &=& \big[\bar u(\bar \alpha)+\Delta U_n \big]^{-1} [\bar v(\bar \alpha)+\Delta V_n \big]
\\
&=& \bar u(\bar \alpha)^{-1} \bar v(\bar \alpha) - \bar u(\bar \alpha)^{-1} \Delta U_n \bar u(\bar \alpha)^{-1} \bar v(\bar \alpha) + \bar u(\bar \alpha)^{-1} \Delta V_n + O(\Delta U_n^2, \Delta U_n \Delta V_n)
\end{eqnarray*}
Using equation \eqref{eq:fixedpoint} this becomes $\alpha_n = \bar\alpha + \Delta \alpha_n$ with
$$
\Delta \alpha_n = - \bar u(\bar \alpha)^{-1} \Delta U_n \bar u(\bar \alpha)^{-1} \bar v(\bar \alpha) + \bar u(\bar \alpha)^{-1} \Delta V_n + O(\Delta U_n^2, \Delta U_n \Delta V_n) \rlap{\ .}
$$
Using the equation \eqref{eq:req-uv} for $\Delta U_n$ and $\Delta V_n$ we can rewrite this as a recursion formula
\begin{equation}
\Delta \alpha_n = \frac{z_{n-1} \Delta \alpha_{n-1} + w_n \Delta a_n}{z_n} + O(\Delta U_n^2, \Delta U_n \Delta V_n)
\label{eq:req-alpha}
\end{equation}
with
$$
\Delta a_n = - \bar u(\bar \alpha)^{-1} \Delta u_n \bar u(\bar \alpha)^{-1} \bar v(\bar \alpha) + \bar u(\bar \alpha)^{-1} \Delta v_n  \rlap{\ .}
$$
By extracting $\Delta u_n $ and $\Delta v_n$ from equation \eqref{eq:duv-order1} we obtain
$$
\Delta a_n = \frac{\partial \bar \alpha(\alpha)}{\partial \alpha}\Big\vert_{\alpha=\bar\alpha} \Delta \alpha_{n-1} + \hat \alpha(\varepsilon_n)
$$
with
$$
\frac{\partial \bar \alpha(\alpha)}{\partial \alpha} = \frac{\partial \big[ \bar u(\alpha)^{-1} \bar v(\alpha)\big] }{\partial \alpha} = - \bar u(\alpha)^{-1} \frac{\partial \bar u(\alpha) }{\partial \alpha} \bar u(\alpha)^{-1} \bar v(\alpha) +  \bar u(\alpha)^{-1} \frac{\partial \bar v(\alpha) }{\partial \alpha}
$$
and introducing
$$
\hat \alpha(\varepsilon_n) = - \bar u(\bar\alpha)^{-1} \hat u(\varepsilon_n) \bar u(\bar\alpha)^{-1} \bar v(\bar\alpha) +  \bar u(\bar\alpha)^{-1} \hat v(\varepsilon_n) \rlap{\ .}
$$
Thus the recursion equation \eqref{eq:req-alpha} can be rewritten at the leading order as
$$
\Delta \alpha_n = \frac{z_{n-1} + w_n \frac{\partial \bar \alpha}{\partial \alpha} }{z_n} \Delta \alpha_{n-1} + \frac{w_n}{z_n} \hat \alpha_n \rlap{\ .}
$$
The solution of this recursion is
\begin{equation}
\Delta \alpha_n = G_{1,n} \frac{\partial \bar \alpha}{\partial \alpha} \Delta \alpha_0 + \sum_{k=1}^n  G_{k,n} \frac{w_k}{z_k} \hat \alpha(\varepsilon_k)
\label{eq:delta-alpha}
\end{equation}
with the linear operator
$$
G_{k,n} = \prod_{j=k+1}^n \frac{z_{j-1} + w_j \frac{\partial \bar \alpha}{\partial \alpha} }{z_j} \rlap{\ .}
$$

$G_{k,n}$ can be computed asymptotically in the limit of large $n$ in the following way. We first rewrite it as
$ G_{k,n} = \prod_{j=k+1}^n \frac{z_{j-1}}{z_j} \prod_{j=k+1}^n \left( 1 + \frac{w_j}{z_{j-1}} \frac{\partial \bar \alpha}{\partial \alpha}\right)$.
The first product simplifies to $\frac{z_k}{z_n}$. The second one behaves as $ \prod_{j=k+1}^n \left( 1 + \frac{w_j}{z_{j-1}} \frac{\partial \bar \alpha}{\partial \alpha}\right) \sim \exp\!\left( \sum_{j=k+1}^n \frac{w_j}{z_{j-1}} \frac{\partial \bar \alpha}{\partial \alpha} \right) $.
As $w_j = z_j - z_{j-1}$, we approximate the discrete sum by an integral: $ \sum_{j=k+1}^n \frac{w_j}{z_{j-1}} \sim \int_{z_k}^{z_n} \frac{\ud z}{z} = \ln\left( \frac{z_n}{z_k}\right)$. Then
 $G_{k,n} \sim \frac{z_k}{z_n} \exp\!\left(   \ln\left( \frac{z_n}{z_k}\right)  \frac{\partial \bar \alpha}{\partial \alpha} \right)$. This finally yields to\footnote{More precisely, if the linear operator $\frac{\partial \bar \alpha}{\partial \alpha}$ has norm $A$: $\left\|\frac{\partial \bar \alpha}{\partial \alpha} \right\|  = A$ for some real number $A \geq 1$, we have $\| G_{k,n} \| \leq A \left( \frac{z_k}{z_n} \right)^{1 - A}$. }:
\begin{equation}
G_{k,n} \sim \left( \frac{z_k}{z_n} \right)^{1 - \frac{\partial \bar \alpha}{\partial \alpha}} \rlap{\ .}
\label{Gapprox}
\end{equation}

We denote by $p_i$ the average price computed over all paths of iteration $i$. As $u_i$ and $v_i$, it depends on the regression coefficients $\alpha_{i-1}$ computed in the previous iteration and on the random variables $\varepsilon_i$ from iteration $i$. Which means that $p_i = p(\alpha_{i-1},\varepsilon_i) = \frac{1}{m}\sum_{j=1}^m p(\alpha_{i-1},\varepsilon_i^j)$ for the $m$ paths of the iteration $i$. Similarly to $u_i$ and $v_i$ the average price on iteration $i$ can be written as the sum of its expected value $\bar p$ and a random part $\hat p$ of null expected value:
$$
p_i = p(\alpha_{i-1},\varepsilon_i) = \bar p(\alpha_{i-1}) + \hat p(\alpha_{i-1},\varepsilon_i)
$$
Expanding $\bar p$ and $\hat p$ around $\bar \alpha$ we rewrite $p_i$ as $p_i = \bar p(\bar \alpha) + \Delta p_i$ with
$\Delta p_i = \frac{\partial \bar p}{\partial \alpha} \Delta \alpha_{i-1} + \hat p(\varepsilon_i)\label{eq:p_i}$
up to higher order terms as in \eqref{eq:uv-exp}. The price after $n$ iterations $P_n$ is the average over $p_i$ with weight $\widetilde{w}_i$:
$$
P_n = \frac{1}{\tilde{z}_n} \sum_{i=1}^n \widetilde{w}_i p_i
$$
with $\tilde{z}_n = \sum_{i=1}^n \widetilde{w}_i $. It also can be written as $P_n = \bar p(\bar \alpha) + \Delta P_n$ with $\Delta P_n = \frac{1}{\tilde{z}_n} \sum_{i=1}^n \widetilde{w}_i \Delta p_i$.
Summing $\Delta p_i$ over $i$ with weights $\widetilde{w}_i$ we have
$$
\Delta P_n = \frac{1}{\tilde{z}_n} \Bigg[ \sum_{i=1}^n \widetilde{w}_i \frac{\partial \bar p}{\partial \alpha}  \Delta \alpha_{i-1} + \sum_{i=1}^n \widetilde{w}_i \hat p(\varepsilon_i) \Bigg]
$$
 Plugging the expression for $\Delta \alpha_{i-1}$ given by equation \eqref{eq:delta-alpha} in this equation we get
\begin{multline}
\Delta P_n = \frac{1}{\tilde{z}_n} \Bigg[\widetilde{w}_1 \frac{\partial \bar p}{\partial \alpha}  \Delta \alpha_0 + \sum_{i=2}^n \widetilde{w}_i \frac{\partial \bar p}{\partial \alpha}  G_{1,i-1} \frac{\partial \bar \alpha}{\partial \alpha} \Delta \alpha_0
\\
+ \frac{\partial \bar p}{\partial \alpha} \sum_{i=2}^n \widetilde{w}_i \sum_{k=1}^{i-1}  G_{k,i-1} \frac{w_k}{z_k} \hat \alpha(\varepsilon_k) + \sum_{i=1}^n \widetilde{w}_i \hat p(\varepsilon_i) \Bigg] \ .
\label{eq:delta-pn}
\end{multline}

The first two terms of equation \eqref{eq:delta-pn} are deterministic and control the expected value of the price error:
$$
\EE(\Delta P_n) = \frac{\partial \bar p}{\partial \alpha} \frac{1}{\tilde{z}_n}\Bigg[\widetilde{w}_1 + \sum_{i=2}^n \widetilde{w}_i  G_{1,i-1} \frac{\partial \bar \alpha}{\partial \alpha} \Bigg] \Delta \alpha_0 \rlap{\ .}
$$
Let us assume that asymptotically, $\widetilde{w}_i \sim w_i$ and therefore $\tilde{z}_n \sim z_n$. Using the asymptotic behavior of $G_{1,i-1}$  from \eqref{Gapprox}, we can rewrite the sum in the previous equation as
$ \sum_{i=2}^n \widetilde{w}_i  G_{1,i-1} \frac{\partial \bar \alpha}{\partial \alpha} \sim \int_{z_1}^{z_n} \left(\frac{z_1}{z}\right)^{1-\frac{\partial \bar \alpha}{\partial \alpha}} \frac{\partial \bar \alpha}{\partial \alpha} \ud z = z_1^{1-\frac{\partial \bar \alpha}{\partial \alpha}} \left( z_n^{\frac{\partial \bar \alpha}{\partial \alpha}} - z_1^{\frac{\partial \bar \alpha}{\partial \alpha}}\right)$.
Thus we get
$$
\EE(\Delta P_n) \sim \frac{\partial \bar p}{\partial \alpha} \bigg(\frac{z_1}{z_n}\bigg)^{1-\frac{\partial \bar \alpha}{\partial \alpha}} \Delta \alpha_0 \rlap{\ .}
$$
This converges to zero if the norm of the operator $\frac{\partial \bar \alpha}{\partial \alpha}$ is smaller than 1:
$ A = \left\| \frac{\partial \bar \alpha}{\partial \alpha} \right\| < 1 $. If asymptotically, $w_i \sim 1$ and therefore $z_n \sim n$ then the convergence is in
$$
\EE(\Delta P_n) \propto \frac{1}{n^{1-A}} \rlap{\ .}
$$

We finally consider the two last terms in equation \eqref{eq:delta-pn}. These are random terms with expected values zero and which are responsible for the variance of the price in the Monte Carlo method. We will study how the variance of these contributions to $P_n$ goes to zero as $n$ goes to infinity. Interverting sums over $i$ and $k$ in the first of these terms, and renaming the mute integer $i$ to $k$ in the last one, we have
$$
\Delta P_n - \EE(\Delta P_n) = \frac{1}{\tilde{z}_n} \Bigg[ \frac{\partial \bar p}{\partial \alpha} \sum_{k=1}^{n-1} \sum_{i=k+1}^n \widetilde{w}_i G_{k,i-1} \frac{w_k}{z_k} \hat \alpha(\varepsilon_k) + \sum_{k=1}^n \widetilde{w}_k \hat p(\varepsilon_k) \Bigg] \rlap{\ .}
$$
Let us introduce $H_{k,n} = \frac{1}{z_k} \sum_{i=k+1}^n \widetilde{w}_i G_{k,i-1}$. As above, we have asymptotically $\sum_{i=k+1}^n \widetilde{w}_i G_{k,i-1} \sim \frac{\partial \bar\alpha}{\partial \alpha} ^{-1}\left(z_k^{1-\frac{\partial \bar\alpha}{\partial \alpha}} z_n^\frac{\partial \bar\alpha}{\partial \alpha} - z_k \right)$
and therefore 
\begin{equation}
H_{k,n} \sim \frac{\left(\frac{z_n}{z_k}\right)^\frac{\partial \bar\alpha}{\partial \alpha} - 1}{\frac{\partial \bar\alpha}{\partial \alpha}} \rlap{\ .}
\label{Hkn}
\end{equation}
If the linear operator $\frac{\partial \bar \alpha}{\partial \alpha}$ has norm $A$,  $H_{k,n}$ has an asymptotic bound. Let us call $y = \frac{z_n}{z_k}$ and $z = \frac{\partial \bar\alpha}{\partial \alpha}$ then using the expansion in series we get $\left\|\frac{y^z-1}{z} \right\| \le  \sum_{n=1}^{\infty} \frac{\left| z\right|^{n-1} \left|\ln y\right|^{n}}{n!}  \le  \sum_{n=1}^{\infty} \frac{\left| A\right|^{n-1} \left|\ln y\right|^{n}}{n!} =\frac{y^A-1}{A}$ as $y > 1$ and $z > 0$.
\begin{equation}
\left\| H_{k,n} \right\| \lesssim \frac{\left( \frac{z_n}{z_k}\right)^A-1}{A} \rlap{\ .}
\label{eq:HknBoundary}
\end{equation}
Using $H_{k,n}$ we have
$\Delta P_n - \EE(\Delta P_n) \sim \frac{1}{z_n} \left[ \frac{\partial \bar p}{\partial \alpha} \sum_{k=1}^{n-1} H_{k,n} w_k \hat\alpha(\varepsilon_k) + \sum_{k=1}^n \widetilde{w}_k \hat p(\varepsilon_k) \right]$.
As $\varepsilon_k$ are independent from each other for different $k$, the variance of the price estimation will be a sum of variances for each $k$:
\begin{multline}
\mathrm{Var}(\Delta P_n) \sim \frac{1}{z_n^2}  \Bigg[ \sum_{k=1}^{n-1} \mathrm{Var}\bigg(  \frac{\partial \bar p}{\partial \alpha}H_{k,n} w_k \hat \alpha(\varepsilon_k)\bigg)
+ \sum_{k=1}^n  \mathrm{Var}\big( \widetilde{w}_k \hat p(\varepsilon_k) \big) 
\\
+ 2 \sum_{k=1}^{n-1} \mathrm{Cov}\bigg( \frac{\partial \bar p}{\partial \alpha} H_{k,n} w_k \hat \alpha(\varepsilon_k) , \widetilde{w}_k \hat  p(\varepsilon_k)\bigg) \Bigg] \rlap{\ .}
\label{eq:var-3terms}
\end{multline}
The expected value of $\hat \alpha$ is zero, so we only get the first term of the variance. Also the quantity in the sum are numbers thus we can see them as  $1 \times 1$ matrices and introduce a trace. Finally we use the cyclic property of the trace and also the linearity of trace and expected value.
\begin{eqnarray}
 \frac{1}{z_n^2}  \sum_{k=1}^{n-1} \mathrm{Var}\left(  \frac{\partial \bar p}{\partial \alpha}H_{k,n} w_k \hat \alpha(\varepsilon_k)\right) 
= \frac{1}{z_n^2}\sum_{k=1}^{n-1} w_k^2 \EE\left[  \hat \alpha(\varepsilon_k)\tr H_{k,n}\tr \frac{\partial \bar p}{\partial \alpha}\tr \frac{\partial \bar p}{\partial \alpha} H_{k,n} \hat \alpha(\varepsilon_k) \right] \nonumber \\
= \frac{1}{z_n^2}\sum_{k=1}^{n-1} w_k^2 \EE\left[ \mathrm{Tr}\left( \hat \alpha( \varepsilon_k)\tr H_{k,n}\tr \frac{\partial \bar p}{\partial \alpha}\tr \frac{\partial \bar p}{\partial \alpha} H_{k,n} \hat \alpha(\varepsilon_k)\right) \right] \nonumber \\
= \mathrm{Tr}\left( \frac{1}{z_n^2}\sum_{k=1}^{n-1} w_k^2  H_{k,n}\tr \frac{\partial \bar p}{\partial \alpha}\tr \frac{\partial \bar p}{\partial \alpha} H_{k,n} \EE\left[ \hat \alpha(\varepsilon_k) \hat \alpha(\varepsilon_k)\tr \right] \right)
\label{eq:Var1-trace}
\end{eqnarray}
$\EE\Big[ \hat \alpha(\varepsilon_k) \hat \alpha(\varepsilon_k)\tr \Big]$ is the covariance matrix of the random part of contributions to $\alpha$. It scales as $\frac{1}{m}$ where $m$ was the number of paths in a given iteration. We therefore write it as
\begin{equation}
\EE\Big[ \hat \alpha(\varepsilon_k) \hat \alpha(\varepsilon_k)\tr \Big] = \frac{1}{m}\Sigma_\alpha
\label{eq:sigma-alpha}
\end{equation}
where $\Sigma_\alpha$ is the variance-covariance matrix of individual contributions to $\alpha_n$.
The question is the convergence of quantity
$$\frac{1}{z_n^2}\sum_{k=1}^{n-1} w_k^2  H_{k,n}\tr \frac{\partial \bar p}{\partial \alpha}\tr \frac{\partial \bar p}{\partial \alpha} H_{k,n} \rlap{\ .}$$
We use the asymptotic behavior of $H_{k,n}$ from \eqref{eq:HknBoundary} in order to get asymptotic boundary. With the assumption that $w_i \sim \widetilde{w}_i \sim 1$ for large $i$ and $z_n \sim \tilde{z}_n \sim n$ we have asymptotically
$$
\big\| H_{k,n} \big\| \lesssim \frac{\big( \frac{n}{k}\big)^A-1}{A}
$$
and therefore
\begin{equation}
\Bigg\| \frac{1}{z_n^2}\sum_{k=1}^{n-1} w_k^2  H_{k,n}\tr \frac{\partial \bar p}{\partial \alpha}\tr \frac{\partial \bar p}{\partial \alpha} H_{k,n} \Bigg\|
\lesssim
\frac{1}{n^2} \sum_{k=1}^{n-1} \bigg\| \frac{\partial \bar p}{\partial \alpha} \bigg\|^2 \frac{1}{A^2} \bigg[\Big( \frac{n}{k}\Big)^A-1\bigg]^2
\label{eq:var1ineq}
\end{equation}
Approximating the sum by an integral we have
$$
\sum_{k=1}^{n-1}  \bigg[\Big( \frac{n}{k}\Big)^A\!-1\bigg]^2 \sim \int_1^n \bigg[\Big( \frac{n}{z}\Big)^A\!-1\bigg]^2 \ud z
= \frac{2 A^2 n}{(1-A)(1-2A)} - \frac{n^{2A}}{1-2A} + \frac{2n^A}{1-A} \rlap{\ .}
$$
Asymptotically the dominating term is the term in $n$ if $A<\frac{1}{2}$ or the term in $n^{2A}$ if $A>\frac{1}{2}$. Using the equations  \eqref{eq:Var1-trace},  \eqref{eq:sigma-alpha} and taking into account the $\frac{1}{n^2}$ in \eqref{eq:var1ineq} we have
\begin{equation}
 \frac{1}{z_n^2}  \sum_{k=1}^{n-1} \mathrm{Var}\bigg(  \frac{\partial \bar p}{\partial \alpha}H_{k,n} w_k \hat \alpha(\varepsilon_k)\bigg)
\lesssim \left\{ \begin{array}{ll}
\displaystyle \frac{d}{(1-A)(1-2A)}  \bigg\| \frac{\partial \bar p}{\partial \alpha} \bigg\|^2\| \Sigma_\alpha\| \frac{1}{m\, n} & \displaystyle \quad A<\frac{1}{2}
\\
\displaystyle \frac{d}{A^2(2A-1)} \bigg\| \frac{\partial \bar p}{\partial \alpha} \bigg\|^2 \| \Sigma_\alpha\| \frac{1}{m\, n^{2-2A}} & \displaystyle \quad A>\frac{1}{2}
\end{array} \right.
\label{eq:Var1}
\end{equation}
where $d$ is the total number of regression functions and comes from the trace. The second term of equation \eqref{eq:var-3terms} is the standard Monte Carlo contribution. As $\hat p(\varepsilon_k) = \frac{1}{m} \sum_{j=1}^m \hat p(\varepsilon_k^j)$  we get $\mathrm{Var}\left( \widetilde{w}_k\hat p(\varepsilon_k) \right)  =  \frac{\widetilde{w}_k^2}{m^2 }\sum_{j=1}^m   \mathrm{Var}\left[ \hat p(\varepsilon_k)^2 \right]$. Let us call $\Sigma_p$ the variance of the payoff on one path $\mathrm{Var}\left[ \hat p(\varepsilon_k)^2 \right]$. For $w_i \sim \widetilde{w}_i \sim 1$ and $z_n \sim \tilde{z}_n \sim n$ we therefore get
\begin{equation}
\frac{1}{z_n^2 }\sum_{k=1}^n  \mathrm{Var}\big( \widetilde{w}_k\hat p(\varepsilon_k) \big)  \sim \frac{1}{m\, n} \Sigma_p \rlap{\ .}
\label{eq:Var2}
\end{equation}

Finally the third term in equation \eqref{eq:var-3terms} comes from the covariance between $\hat \alpha$ and $\hat p$. Using linearity of expected value and the fact that both $\hat \alpha$ and $\hat p$ have null expected values by construction, we rewrite it as
\begin{equation}
\frac{2}{z_n^2} \sum_{k=1}^{n-1} \mathrm{Cov}\bigg( \frac{\partial \bar p}{\partial \alpha} H_{k,n} w_k \hat \alpha(\varepsilon_k) , \widetilde{w}_k\hat p(\varepsilon_k)\bigg) = \frac{2}{z_n^2}  \frac{\partial \bar p}{\partial \alpha} \sum_{k=1}^{n-1} H_{k,n} w_k \widetilde{w}_k \EE\Big[ \hat \alpha(\varepsilon_k) \hat p(\varepsilon_k) \Big] \rlap{\ .}
\label{eq:Var3-eq}
\end{equation}
Similarly to the first terms, we can write $\EE\Big[ \hat \alpha(\varepsilon_k) \hat p(\varepsilon_k) \Big] = \frac{1}{m} \Sigma_{\alpha p}
$
where $\Sigma_{\alpha p}$ is the covariance between individual path contributions to regression coefficients $\alpha$ and price $p$. With $\widetilde{w}_i \sim w_i \sim 1$ and $\tilde{z}_n \sim z_n \sim 1$ and using asymptotic expression \eqref{Hkn} for $H_{k,n}$ the sum over $k$ in equation \eqref{eq:Var3-eq} becomes asymptotically
$$
\frac{1}{z_n^2} \sum_{k=1}^{n-1} H_{k,n} w_k \widetilde{w}_k \sim \frac{1}{n^2}\sum_{k=1}^{n-1} \frac{\big(\frac{n}{k}\big)^\frac{\partial \bar\alpha}{\partial \alpha} - 1}{\frac{\partial \bar\alpha}{\partial \alpha}}
\rlap{\ .}
$$
Approximating the discrete sum by an integral, this gives
$$
\frac{1}{z_n^2} \sum_{k=1}^{n-1} H_{k,n} w_k \widetilde{w}_k \sim \frac{1}{n^2}\int_1^n \frac{\big(\frac{n}{z}\big)^\frac{\partial \bar\alpha}{\partial \alpha} - 1}{\frac{\partial \bar\alpha}{\partial \alpha}} \ud z = \frac{1}{1-\frac{\partial \bar\alpha}{\partial \alpha}} \frac{1}{n} - \frac{1}{\frac{\partial \bar\alpha}{\partial \alpha} \Big( 1-\frac{\partial \bar\alpha}{\partial \alpha}\Big)} \frac{1}{n^{2-\frac{\partial \bar\alpha}{\partial \alpha}}} + \frac{1}{\frac{\partial \bar\alpha}{\partial \alpha}} \frac{1}{n^2}
\rlap{\ .}
$$
For $A = \big\| \frac{\partial \bar\alpha}{\partial \alpha} \big\| < 1$ the leading term is the first one:
$$
\frac{1}{z_n^2} \sum_{k=1}^{n-1} H_{k,n} w_k \widetilde{w}_k \sim \frac{1}{1-\frac{\partial \bar\alpha}{\partial \alpha}} \frac{1}{n} \rlap{\ .}
$$
We thus have
\begin{equation}
\frac{2}{z_n^2} \sum_{k=1}^{n-1} \mathrm{Cov}\bigg( \frac{\partial \bar p}{\partial \alpha} H_{k,n} w_k \hat \alpha(\varepsilon_k) , \widetilde{w}_k\hat p(\varepsilon_k)\bigg)
\sim
\frac{2}{1-\frac{\partial \bar\alpha}{\partial \alpha}} \Sigma_{\alpha p} \frac{1}{m\, n} \rlap{\ .}
\label{eq:Var3}
\end{equation}

Summing the terms \eqref{eq:Var1}, \eqref{eq:Var2} and \eqref{eq:Var3} we finally find that the variance of the price behaves as
$$
\mathrm{Var}(\Delta P_n) \propto \frac{1}{m \,n^{\min(1,2-2A)}}
$$
which gives a standard error in
$$
\sqrt{\mathrm{Var}(\Delta P_n)} \propto \frac{1}{\sqrt{m} \, n^{\min\big(\frac{1}{2},1-1A\big)}}
$$

We finally obtained that the expected value of the error decreases in $\frac{1}{\sqrt{m}\, n^{1-A}}$ and that the statistical error decreases with a power given by the minimum of the same $\frac{1}{\sqrt{m}\, n^{1-A}}$ and the usual Monte Carlo error in $\frac{1}{\sqrt{m\, n}} = \frac{1}{\sqrt{N}}$.
\end{proof}
\end{appendices}

\subsection*{Acknowledgments}
We thank Julie Barth\'es, Sergey Derzho, Nicholas Leib, Martial Millet and Arnaud Rivoira for useful comments.

\bibliographystyle{apalike}
\bibliography{AMC}

\end{document}